\documentclass[12pt]{article}
\usepackage{amssymb,amsmath,amscd}
\usepackage{epsfig}
\usepackage{epstopdf}
\usepackage{latexsym}
\usepackage{graphicx}
\usepackage{booktabs}
\usepackage{bbm}
\usepackage[margin=15pt,small]{caption}
\usepackage{subcaption}
\usepackage{enumitem}
\usepackage{float}
\usepackage[toc]{appendix}
\usepackage[numbers,compress]{natbib}
\usepackage{array}
\usepackage{autobreak}
\usepackage{listings}
\usepackage{longtable}
\allowdisplaybreaks
\usepackage{color}

\definecolor{cardinal}{rgb}{0.6,0,0}
\definecolor{darkgreen}{rgb}{0.1,0.4,0}
\definecolor{golden}{rgb}{0.92, 0.7, 0}
\definecolor{midnight}{rgb}{0, 0, 0.5}
\definecolor{darkblue}{rgb}{0.3,0.3,0.7}
\definecolor{darkred}{rgb}{0.7,0,0}

\usepackage[T1]{fontenc}
\usepackage{lmodern}
\usepackage{datetime}

\usepackage{hyperref}
\hypersetup{
  colorlinks=true,
  linkcolor=darkblue,
  citecolor=darkred,
  urlcolor=darkgreen,
  linktocpage=true,
  linktoc=page
}

\setlist{nolistsep}

\numberwithin{equation}{section} 
\numberwithin{table}{section}
\numberwithin{figure}{section}

\newcolumntype{L}{>{$}l<{$}}
\allowdisplaybreaks

\let\oldbibliography\thebibliography
\renewcommand{\thebibliography}[1]{\oldbibliography{#1}
\setlength{\itemsep}{0pt}} 

\def\cN{{\cal N}}
\def\diag{{\rm diag}}

\newcommand{\Pu}[1]{{\tt P#1}}

\newcommand{\be}{\begin{equation}}
\newcommand{\ee}{\end{equation}}
\newcommand{\bea}{\begin{eqnarray}}
\newcommand{\eea}{\end{eqnarray}}

\newcommand{\f}[2]{\frac{#1}{#2}}
\newcommand{\R}{\mathbb{R}}
\newcommand{\Z}{\mathbb{Z}}
\newcommand{\p}[1]{\phantom{#1}}

\newcommand{\Tr}{\text{Tr}~}

\newcommand{\U}{\text{U}}
\newcommand{\SU}{\text{SU}}
\newcommand{\SO}{\text{SO}}
\newcommand{\OO}{\text{O}}
\newcommand{\ISO}{\text{ISO}}

\newcommand{\SL}{\text{SL}}

\newcommand{\su}{\mathfrak{su}}

\newcommand{\ssl}{\mathfrak{sl}}
\newcommand{\Ess}{\text{E}_{7(7)}}

\newcommand{\ess}{\mathfrak{e}_{7(7)}}

\def\eql{~=~}
\def\coeff#1#2{\relax{\textstyle \frac{#1}{#2}}\displaystyle}

\newcommand{\bb}[1]{\mathbb{#1}}

\begin{document}  

\pagenumbering{Alph}
\begin{titlepage}
 
\medskip
\begin{center} 
{\Large \bf New AdS$_4$ Vacua in Dyonic ISO(7) Gauged Supergravity} 

\bigskip
\bigskip
\bigskip
\bigskip

{\bf Nikolay Bobev,${}^{\rm A}$ Thomas Fischbacher,${}^{\rm d}$ \\Fri\dh rik Freyr Gautason,${}^{\rm S}$ and Krzysztof Pilch${}^{4}$   \\ }
\bigskip
${}^{\rm A}$
Instituut voor Theoretische Fysica, KU Leuven,\\ 
Celestijnenlaan 200D, B-3001 Leuven, Belgium
\vskip 5mm
${}^{\rm d}$ Google Research\\
Brandschenkestrasse 110, 8002 Z\"urich, Switzerland
\vskip 5mm
${}^{\rm S}$ University of Iceland, Science Institute\\
Dunhaga 3, 107 Reykjav\'ik, Iceland
\vskip 5mm
${}^{4}$ Department of Physics and Astronomy \\
University of Southern California \\
Los Angeles, CA 90089, USA  \\
\bigskip
\tt{nikolay.bobev@kuleuven.be, ffg@hi.is, tfish@google.com, pilch@usc.edu}  \\
\end{center}

\bigskip
\bigskip

\begin{abstract} \noindent  We identify 219 AdS$_4$ solutions in four-dimensional dyonically gauged $\rm ISO(7)$ $\mathcal{N}=8$ supergravity and present some of their properties. One of the new solutions preserves $\mathcal{N}=1$ supersymmetry and provides a rare explicit example of an AdS$_4$ vacuum dual to a 3d SCFT with no continuous global symmetry. There are also two new non-supersymmetric solutions for which all 70 scalar fields in the supergravity theory have masses above the BF bound. All of these AdS$_4$ solutions can be uplifted to massive type IIA supergravity. Motivated by this we present the low lying operator spectra of the dual 3d CFTs for all known supersymmetric AdS$_4$ solutions in the theory and organize them into superconformal multiplets.

\end{abstract}
\end{titlepage}
\pagenumbering{arabic}

\newpage
\setcounter{tocdepth}{2}
\tableofcontents

\section{Introduction and outlook}
\label{sec:Introduction}

Charting the terra incognita of consistent AdS$_4$ vacua is a challenging but important task for the explorers of the string theory landscape. A valuable collection of such explicit vacua arises from consistent truncations of 10d or 11d supergravity to four-dimensional gauged supergravity. Our goal in this paper is to describe one such corner of the landscape of AdS$_4$ vacua in massive IIA supergravity \cite{Romans:1985tz}.

In our search for AdS$_4$ vacua, we explore a 4d $\mathcal{N}=8$ supergravity theory with a specific dyonic $\ISO(7)$ gauging \cite{DallAgata:2014tph,Guarino:2015qaa}. This theory arises as a consistent truncation of massive type IIA supergravity on $S^6$ and, in particular, there are explicit uplift formulae that allow to map every 4d solution to a solution in the 10d theory \cite{Guarino:2015vca}. A notable feature of the 4d gauged supergravity is the non-trivial potential for the 70 scalar fields in the theory. The question of finding interesting AdS$_4$ vacua then translates into finding critical points of this potential. Unfortunately, due to its algebraic complexity,  it is hard to compute the potential in closed analytic form. 

In this paper, we employ two different methods that have been developed to search for critical points in this context. The first one, proposed by Warner in \cite{Warner:1983vz},  amounts to imposing invariance under a suitable subgroup of the symmetry group of the supergravity theory. This results in a consistent truncation to a theory with a smaller number of scalar fields in which the scalar potential can be computed and extremized analytically. The second approach is a  full numerical search for the critical points of the scalar potential as a function of all scalar fields. The latter approach has been used in a series of papers to explore the scalar potentials of 3d and 4d maximal gauged supergravity \cite{Fischbacher:2002fx,Fischbacher:2008zu,Fischbacher:2009cj,Fischbacher:2011jx,Fischbacher:2010ki}
 and led to the application of machine learning techniques based on Google's \texttt{TensorFlow} platform  \cite{Abadi2016} to the 70-scalar potential of the 4d $\mathcal{N}=8$ de Wit-Nicolai $\SO(8)$ gauged supergravity theory \cite{deWit:1982bul} in  \cite{Comsa:2019rcz} (see, also  \cite{Bobev:2019dik}).\footnote{The same techniques were also used in  \cite{Krishnan:2020sfg,Bobev:2020ttg}  to find AdS$_5$ vacua in the 5d $\mathcal{N}=8$ $\SO(6)$ gauged supergravity.} 

Our approach is effectively a mixture of the two methods.\footnote{There is also a third method used to search for AdS$_4$ vacua in gauged supergravity which is based on the embedding tensor formalism, see \cite{Dibitetto:2011gm,DallAgata:2011aa} and references thereof. We do not use this method in the present work.} We first identify a suitable $\mathbb{Z}_2\times \mathbb{Z}_2$ symmetry subgroup  and use it to truncate the full 4d $\mathcal{N}=8$ supergravity to the invariant sector which is 4d $\mathcal{N}=2$ gauged supergravity with 3 vector multiplets and 4 hypermultiplets. This model has 22 real scalar fields and the supergravity potential can be computed analytically by employing the so-called ``solvable parametrization'' based on the Iwasawa decomposition of the scalar coset manifold. Unfortunately,  the closed form analytic expression for the potential, which we compute explicitly, is  quite unwieldy and finding its critical points analytically is still  prohibitively difficult. Hence we turn to numerical routines based on \texttt{Mathematica}  to perform a systematic search for critical points. This results in 219 distinct AdS$_4$ vacua that we find. We have made extensive crosschecks of our results against an ongoing comprehensive search for AdS$_4$ vacua based on Google's \texttt{TensorFlow}  of the full supergravity potential in \eqref{eq:4dN8potfull} \cite{BFGP}.

There is also a smaller  consistent truncation of the dyonically ISO(7) gauged supergravity obtained by imposing an additional $\mathbb{Z}_2$ symmetry,  which was considered previously in \cite{Guarino:2020jwv}. It consists of 
 4d $\mathcal{N}=1$ supergravity coupled to 7 chiral multiplets. The 14-scalar potential in this  truncation can also be studied explicitly and we find that it has 65 of the 219 critical points of the 22-scalar model. Most of these 65 points have been found previously. As summarized in \cite{Guarino:2020jwv}, 60 AdS$_4$ vacua have already been found in previous studies of the 14-scalar model and its sub-truncations, see also \cite{Guarino:2015jca,Guarino:2016ynd,Guarino:2019jef,Guarino:2019snw}. The net result of our full search is therefore the identification of 159 new critical points of the dyonically gauged ISO(7) supergravity theory.

There are 7 supersymmetric critical points in the full list of 219 points. One of these supersymmetric points is new. In addition, there are 9 non-supersymmetric AdS$_4$ solutions, 2~of which are new, for which all 70 scalars of the 4d $\mathcal{N}=8$ supergravity have masses above the BF~bound \cite{Breitenlohner:1982jf}  and thus   are perturbatively stable within the 4d ${\cal  N}=8$ theory. We present the spectra of mass fluctuations for all bosonic and fermionic 4d $\mathcal{N}=8$ supergravity fields around each of these 16 perturbatively stable solutions. For the 7 supersymmetric critical points we organize these spectra into supersymmetric multiplets and map them to the spectrum of operators in the dual 3d SCFT. The new supersymmetric solution is of special interest. It preserves $\mathcal{N}=1$ supersymmetry and does not exhibit any continuous symmetry. Uplifting this solution to massive IIA supergravity using the formulae in \cite{Guarino:2015vca} will provide a very rare example of a fully explicit stable AdS$_4$ solution of string theory which does not have any continuous internal symmetry.\footnote{The only other explicit example we are aware of is the AdS$_4$ $\mathcal{N}=1$ J-fold solution found recently in \cite{Bobev:2019jbi}. To ensure no isometry in the internal space one needs to choose an appropriate K\"ahler-Einstein manifold for the IIB solution in \cite{Bobev:2019jbi}.} Using holography, this implies that there is a new 3d $\mathcal{N}=1$ SCFTs arising on the worldvolume of D2-branes, which has no conserved flavor currents and only a discrete global symmetry.

Our results suggest several directions for further study some of which we would like to mention briefly now before moving on to the technical part of this paper. 

It is now clear that the full scalar potential of the 4d $\mathcal{N}=8$ dyonic ISO(7) supergravity with all 70 scalars  should have a very large number of critical points corresponding to a plethora of AdS$_4$ solutions of the theory.   Compiling a catalog of   these  points  and their properties will be helpful to better understand the structure of the theory.   We plan to report further results in this direction  in~\cite{BFGP}.

The 7 supersymmetric AdS$_4$ solutions that we have identified are necessarily non-perturbatively stable, see~\cite{Gibbons:1983aq}. Analyzing the full stability of the 9 non-supersymmetric BF stable solutions that are now known is more subtle.  It was shown in \cite{Guarino:2020jwv} that 7 of these solutions, the ones that belong to the 14-scalar model, do not suffer from the Brane-Jet instability discussed in \cite{Bena:2020xxb}. Similar analysis should be carried out for the  2 new BF stable non-supersymmetric critical points. Even more interesting is to understand the perturbative stability of these points within the massive IIA supergravity using recent methods for computing the Kaluza-Klein mass spectra   developed in \cite{Malek:2019eaz,Malek:2020yue}. Indeed, as shown very recently in  \cite{Guarino:2020flh}, one of the previously found non-supersymmetric critical points with a G$_2$ symmetry appears to be stable. It would be interesting to understand what happens with the 2 new non-supersymmetric BF stable points we find here. In view  of the AdS Swampland Conjecture \cite{Ooguri:2016pdq}, one might expect some non-perturbative mechanism triggering an instability that is yet to be found.

Our results should also have implications for the physics of 3d SCFTs arising from D2-branes in massive IIA string theory. The 3d $\mathcal{N}=8$ SYM theory on the worldvolume of the branes is modified by the presence of a Chern-Simons term induced by the  Romans mass. In addition, one can have superpotential mass terms in  this theory, which corresponds to turning on background fluxes and metric deformations in the type IIA string theory. Understanding the ensuing RG flows and the low-energy phase diagram of this class of 3d QFTs is in general a complicated problem even when some supersymmetry is preserved. Holography is an indispensable tool to understand this physics. The existence of the 7 supersymmetric AdS$_4$ solutions discussed above suggests that there is a rich web of IR SCFTs  connected by RG flows. Some of these RG flows were studied in \cite{Guarino:2016ynd} and the dual CFTs for two of the supersymmetric critical points preserving $\mathcal{N}=2$ and $\mathcal{N}=3$ supersymmetry were identified in \cite{Guarino:2015jca} and \cite{Guarino:2019snw}, respectively. However, to understand the physics of the 3d $\mathcal{N}=1$ SCFTs dual to the 5 other supersymmetric critical points is far more challenging. This will certainly be the case for  the 3d $\mathcal{N}=1$ SCFT dual to the new AdS$_4$ vacuum that we identify in this paper. Using the spectrum of operators with low conformal dimensions we compute here, together with the $\Z_2\times \Z_2$ global symmetry, should facilitate the study of this problem, perhaps by employing the techniques of the $\mathcal{N}=1$ superconformal bootstrap \cite{Bashkirov:2013vya,Rong:2018okz,Atanasov:2018kqw}. 

Finally, we should emphasize that the results of this search exhibit  the familiar predicament known from the previous searches (see, e.g., \cite{Comsa:2019rcz,Bobev:2019dik,Bobev:2020ttg}) that most of the critical points in the maximal gauged supergravity theories arising from string and M-theory are non-supersymmetric and BF unstable. As discussed in \cite{Kaplan:2009kr}, such instabilities of AdS vacua signal a loss of conformal invariance in the dual QFT. Given that all these unstable AdS solutions arise from string theory and appear to belong to rich webs of interconnected RG flows, it is certainly desirable to understand better their physics and the implications for the dual QFT.

The paper is organized as follows. We continue our discussion in Section~\ref{sec:sugra} with a short summary of the salient features of the dyonic ISO(7) gauged supergravity and describe our choice of parametrization of the scalar manifold, delegating some of the details to Appendix~\ref{sec:roots}. In Section~\ref{sec:vacua} we present the result of the numerical search for AdS$_4$ solution in this model. In Section~\ref{sec:new} we identify the 16 critical points in our list for which all 70 scalar masses obey the BF bound. We also present the mass spectra of the new $\mathcal{N}=1$ AdS$_4$ solution and the corresponding spectrum of operators in the dual 3d SCFT. In Appendix~\ref{sec:sysyAdS4vacua} we present the mass spectra of the 6 other supersymmetric vacua we find in this model and show how to organize the operator dimensions in the dual SCFTs into superconformal multiplets. The mass spectra for the 9 non-supersymmetric BF stable solutions are presented in Appendix~\ref{sec:nonsusy}.

\section{Dyonic ISO(7) gauged supergravity}
\label{sec:sugra}

In this section we give a brief overview of the dyonic $\ISO(7)$ gauged supergravity in four dimensions. We focus on the structure of the scalar potential in this model. Further details of the full gauged supergravity theory are given in \cite{Guarino:2015qaa} using the embedding tensor formalism of \cite{deWit:2005ub,deWit:2007kvg}. As shown in \cite{Guarino:2015jca,Guarino:2015vca}, this 4d gauged supergravity arises as a consistent truncation of massive type IIA supergravity \cite{Romans:1985tz} on $S^6$ .

The 70 scalar fields of ${\cal N}=8$ supergravity in four dimensions parametrize the coset manifold $\Ess/(\SU(8)/\mathbb{Z}_2)$. To obtain the scalar potential in the theory we need to calculate  the vielbein ${\cal V}$ on the coset that depends on the 70 scalar fields. We start by constructing the 133 basis elements of $\ess$ written as $56 \times 56$ matrices.  We use the real $\ssl(8)$ basis for $\ess$ (see for example \cite{Cremmer:1979up}) where the infinitesimal transformation acts on a pair of $28$-dimensional vectors written as antisymmetric tensors $x_{[AB]}$ and $x^{[AB]}$ with $A,B=1,\cdots,8$. These tensors can be combined into a single ${\bf 56}$-vector $x_{\bb M} = (x_{[AB]},x^{[AB]})$
\be\label{Essgenerator}
\begin{split}
\delta x_{[AB]} &= \Lambda_{A}{}^C x_{[CB]} + \Lambda_B{}^Cx_{[AC]} + \Sigma_{ABCD} x^{[CD]}\,,\\
\delta x^{[AB]} &= -\Lambda_C{}^{A}x^{[CB]} - \Lambda_C{}^{B}x^{[AC]} + \Sigma^{ABCD} x_{[CD]}\,.
\end{split}
\ee
Here $\Lambda_A{}^{B}$ is a traceless $\ssl(8)$ matrix, $\Sigma_{ABCD}$ is totally antisymmetric, and $\Sigma^{ABCD}$ is its dual
\be
\Sigma^{ABCD} = \f{1}{4!}\epsilon^{ABCDEFGH} \Sigma_{EFGH}\,. 
\ee
The transformation parameters in \eqref{Essgenerator} have been split into $(\Lambda^A_{\p{A}B},\Sigma_{ABCD})$ according to the branching rule ${\bf 133} = {\bf 63} \oplus {\bf 70}$ of the adjoint representation of $\Ess$ under $\SL(8,\R)$. This can be used to write a general element of $\ess$ as
\be
\mathfrak{X}_{\bb M}{}^{\bb N}= \left(\begin{matrix} 2\Lambda_{[A}^{\p{[A}[C}\delta_{B]}^{D]} & \Sigma_{[AB][CD]}\\ \Sigma^{[AB][CD]} & -2\Lambda_{[C}^{\p{[C}[A}\delta^{B]}_{D]} \end{matrix}\right)\,.
\ee
One then finds the following result for the $\ess$ Killing form 
\be
\Tr(\mathfrak{X}_1\cdot \mathfrak{X}_2) = 12 \Tr(\Lambda_1 \cdot \Lambda_2) + 2 \Sigma_{1\,ABCD}\Sigma_2^{ABCD}\,.
\ee
In particular, this shows that the 70 non-compact generators are obtained by selecting $\Lambda$ symmetric and $\Sigma_{ABCD}$ self-dual. The 63 remaining generators are obtained by specifying $\Lambda$ antisymmetric and $\Sigma_{ABCD}$ anti-self-dual form the $\su(8)$ subalgebra of $\ess$.

We will work in a basis of generator for $\ess$ obtained by choosing a suitable basis for $\Lambda^A_{\p{A}B}$ and $\Sigma_{ABCD}$, namely 
\be
\begin{split}
t_A^{\p{A}B}~&:\quad \Lambda_C{}^D = \tfrac{1}{\sqrt{2}}\delta_C^B\delta_A^D\,,\quad \Sigma=0\,, \\
t_{ABCD}~&:\quad  \Sigma_{EFGH} = \tfrac{1}{2\sqrt{2}}\epsilon_{ABCDEFGH}\,,\quad \Lambda=0\,,
\end{split}
\ee
where the normalization coefficients have been chosen for convenience. Notice that the diagonal $t_A{}^B$ generators do not correspond to traceless $\Lambda$. Indeed we have eight $t_A{}^A$ which should always be combined in such a way so that the trace is removed. We will come back to this below.

\subsection{Solvable parametrization}
\label{subsec:solvable}

After defining a basis for the generators $\Ess$ we are ready to discuss the parametrization of the scalar manifold. We employ the so-called solvable parametrization \cite{Andrianopoli:1996bq,Andrianopoli:1996zg}, in which the scalar vielbein is given by
\be
{\cal V} = \exp\left(\varphi_n \mathfrak{h}_n\right)\cdot \exp\Big(\sum_{\alpha\in\Delta_+}\theta_\alpha \mathfrak{e}_\alpha\Big)\,,
\ee
where $\mathfrak{h}_i$ are the generators of a noncompact Cartan subalgebra and $\mathfrak{e}_\alpha$ are generators of a nilpotent subalgebra corresponding to positive root generators. This parametrization has several advantages; firstly the exponents are relatively simple to compute, and secondly the truncation  with respect to  discrete symmetries discussed below is straightforward.

To proceed we select a noncompact Cartan subalgebra defined by a combination of diagonal generators in $\ssl(8,\R)$. We need to combine them appropriately to obtain a proper traceless $\ssl(8,\R)$ matrix as discussed above. This results in seven Cartan generators given by
\begin{equation}\label{cartangenerators}
\mathfrak{h}_{n} = \frac{1}{\sqrt{n(n+1)}}\Big(\sum_{i=1}^n t_i^{\p{i}i}-n \,t_{n+1}^{\p{n+1}n+1}\Big)\,, \qquad n=1,\ldots 7\,.
\end{equation}
The corresponding  positive and negative root generators, $\mathfrak{e}_\alpha$ and $\mathfrak{f}_\alpha$, respectively, are given in Table~\ref{tbl:roots} in Appendix \ref{sec:roots}. They obey the following identities
\be
[\mathfrak{h}_n,\mathfrak{e_\alpha}] = \alpha_n \mathfrak{e_\alpha}\,,\qquad [\mathfrak{h}_n,\mathfrak{f_\alpha}] = -\alpha_n \mathfrak{f_\alpha}\,.
\ee
We parametrize the positive roots, $\boldsymbol{\alpha}=(\alpha_1,\alpha_2,\dots,\alpha_7)\in\Delta_+$, in terms of their coordinates in the simple root basis,
\begin{equation}\label{rootnot}
[n_1n_2n_3n_4n_5n_6n_7]\qquad \longleftrightarrow\qquad \boldsymbol{\alpha}=\sum_{i=1}^7 n_i\boldsymbol{\alpha}_{(i)}\,,
\end{equation}
where the simple roots, $\boldsymbol{\alpha}_{(i)}$, are given explicitly by 
\begin{equation}\label{}
\begin{aligned}
\boldsymbol{\alpha}_{(1)}   &\eql  \left(\sqrt{2},0,0,0,0,0,0\right)\,, &  
\boldsymbol{\alpha}_{(2)}   &\eql \left(-\coeff{1}{\sqrt{2}},\sqrt{\coeff{3}{2}},0,0,0,0,0\right)\,,\\ 
\boldsymbol{\alpha}_{(3)}   &\eql \left(0,-\sqrt{\coeff{2}{3}},\coeff{2}{\sqrt{3}},0,0,0,0\right)\,,&\qquad 
\boldsymbol{\alpha}_{(4)}   &\eql \left(0,0,-\coeff{\sqrt{3}}{2},\coeff{\sqrt{5}}{2},0,0,0\right)\,,\\
\boldsymbol{\alpha}_{(5)}   &\eql  \left(0,0,0,-\coeff{2}{\sqrt{5}},\sqrt{\coeff{6}{5}},0,0\right)\,,&    
\boldsymbol{\alpha}_{(6)}   &\eql  \left(0,0,0,0,-\sqrt{\coeff{5}{6}},\sqrt{\coeff{7}{6}},0\right)\,,\\
\boldsymbol{\alpha}_{(7)}   &\eql  \left(0,0,0,-\coeff{2}{\sqrt{5}},-\sqrt{\coeff{8}{15}},-\sqrt{\coeff{8}{21}},-\sqrt{\coeff{2}{7}}\right)\,.
\end{aligned}
\end{equation}
The normalization above are chosen such that
\begin{equation}\label{}
\Tr \mathfrak{h}_m\mathfrak{h}_n=12\,\delta_{mn}\,, \qquad \Tr \mathfrak{e}_\alpha \mathfrak{e}_\beta\eql \Tr \mathfrak{f}_\alpha \mathfrak{f}_\beta\eql 0\,,\qquad \Tr \mathfrak{e}_\alpha \mathfrak{f}_\beta\eql 6\,\delta_{\alpha\beta}\,. 
\end{equation}
%

\subsection{The scalar potential}

With the coset parametrization at hand we move on to some of the relevant details of the supergravity theory. We are interested in searching for AdS$_4$ vacua and thus we need to study the critical points of the scalar potential $V$. To this end we focus on the scalar part of the supergravity Lagrangian which reads
\be
{\cal L} = \sqrt{-g}\left( R + \f{1}{96}D_\mu {\cal M}^{\bb M \bb N} D^\mu {\cal M}_{\bb M\bb N}-V\right)\,,
\ee
where
\be
{\cal M} = {\cal V}\cdot {\cal V}^T\,,\quad {\cal M}^{\bb M \bb R} {\cal M}_{\bb R\bb N} = \delta_{\bb N}^{\bb M}\,,
\ee
and $D_\mu$ is the gauge covariant derivative. The scalar potential is given by \cite{deWit:2007kvg,Guarino:2015qaa}
\be\label{eq:4dN8potfull}
V = \f{g^2}{672}{\cal M}^{\bb M\bb P}X_{\bb M\bb N}^{\p{\bb M\bb N}\bb R} X_{\bb P\bb Q}^{\p{\bb P\bb Q}\bb S} \left({\cal M}^{\bb N\bb Q}{\cal M}_{\bb R\bb S} + 7 \delta_{\bb S}^{\bb N} \delta_{\bb R}^{\bb Q} \right)\,,
\ee
where $g$ is the gauge coupling constant. For the $\ISO(7)$ gauging of the $\mathcal{N}=8$ theory the tensor $X$ can be obtained from the embedding tensor specifying the gauging and is given by \cite{Guarino:2015qaa}
\be
\begin{split}
X_{[AB][CD]}{}^{[EF]} &= -X_{[AB]}{}^{[EF]}{}_{[CD]} = - 8 \delta^{[E}_{[A}\theta_{B][C}\delta_{D]}^{F]}\,,\\
X^{[AB]}{}_{[CD]}{}^{[EF]} &= -X^{[AB][EF]}{}_{[CD]} = - 8 \delta^{[A}_{[C}\xi^{B][E}\delta_{D]}^{F]}\,,
\end{split}
\ee
where
\be
\theta = \diag(1,1,1,1,1,1,1,0)\,,\qquad \xi = \diag(0,0,0,0,0,0,0,c)\,.
\ee
The parameter $c$ characterizes the type of $\ISO(7)$ gauging. It was shown in \cite{DallAgata:2014tph} that there are two inequivalent gaugings. For $c=0$ the gauging is purely electric and one recovers the $\ISO(7)$ gauged supergravity theory constructed by Hull in \cite{Hull:1984yy}. This 4d $\mathcal{N}=8$ theory arises as a consistent truncation of the type IIA supergravity with vanishing Romans mass on $S^6$.  For $c\ne0$ the seven translations inside $\ISO(7)$ are dyonically gauged with a coupling constant $m = g c$. As discussed in \cite{DallAgata:2014tph} for all values of $c\neq 0$ the theory is equivalent and can be obtained by a consistent truncation of the massive IIA supergravity on $S^6$ \cite{Guarino:2015jca,Guarino:2015vca}.\footnote{The parameter $m$ is proportional to the Romans mass in the massive IIA supergravity.} From now on we work in conventions where we set $c=1$ and also set the parameter $g=1$ which in turn fixes the AdS$_4$ length scale.

\subsection{Discrete symmetries and truncations}

Since the full scalar potential is difficult to compute as a function of the 70 scalar fields, we focus on truncations using three $\Z_2\subset\SL(8,\R)$ symmetries that act on an eight-dimensional vector as follows \cite{Guarino:2019jef}: 
\be\label{Sdef}
\begin{split}
S_1\,:~(x_1,x_2,x_3,x_4,x_5,x_6,x_7,x_8)&\mapsto (x_1,x_2,x_3,-x_4,-x_5,-x_6,-x_7,x_8)\,,\\
S_2\,:~(x_1,x_2,x_3,x_4,x_5,x_6,x_7,x_8)&\mapsto (x_1,-x_2,-x_3,x_4,x_5,-x_6,-x_7,x_8)\,,\\
S_3\,:~(x_1,x_2,x_3,x_4,x_5,x_6,x_7,x_8)&\mapsto (x_1,-x_2,x_3,-x_4,x_5,-x_6,x_7,-x_8)\,.
\end{split}
\ee
One can impose invariance under these $\mathbb{Z}_2$ symmetries in order to construct consistent truncations of the 4d $\mathcal{N}=8$ gauged supergravity. Three inequivalent truncation may be obtained by keeping fields invariant under one, two, or all three actions. We now proceed to discuss three different truncations obtained in this way.

Keeping the supergravity fields that are invariant under one of the three discrete symmetries in \eqref{Sdef}, say $S_1$, results in an ${\cal N}=4$ supergravity theory, see \cite{Dibitetto:2011eu,Guarino:2019jef}. The invariant fields include the metric as well as 38 real scalars that parametrize the manifold
\be
\f{\SO(6,6)}{\SO(6)\times\SO(6)} \times \f{\SL(2,\R)}{\U(1)}\,.
\ee
In addition, one finds 12 invariant vector fields that transform in the adjoint of $\SO(4)\times \ISO(3)$. The invariant fermions comprise of 4 gravitini and 28 gaugini. These fields can indeed be organized into an $\mathcal{N}=4$ gravity multiplet and 6 vector multiplets to form a full  $\mathcal{N}=4$ supergravity theory, see for instance \cite{Schon:2006kz}. Unfortunately it is prohibitively hard to compute explicitly the potential in this truncation analytically. 

A more tractable truncation can be obtained by imposing invariance with respect to two of the three $\Z_2$ actions \eqref{Sdef}, say $S_1$ and $S_2$. The invariant fields consist of the metric together with 22 real scalar fields parametrizing the manifold
\be\label{eq:22scman}
\f{\SO(4,4)}{\SO(4)\times\SO(4)} \times \Bigg[\f{\SL(2,\R)}{\U(1)}\Bigg]^3\,.
\ee
In addition one also finds that $\OO(1,1)\times \SO(2)^3$ gauge fields, 2 gravitini and 14 gaugini are left invariant. These fields can be organized into a full 4d $\mathcal{N}=2$ supergravity theory consisting of the $\mathcal{N}=2$ gravity multiplet coupled to 3 vector multiplets and 4 hypermultiplets. We were able to find an analytic expression for the scalar potential in this model and in the next section we perform a systematic search and find 219 critical points. Since the generators $S_1$ and $S_2$ generate the discrete group $\Z_2\times \Z_2$ (which is isomorphic to the Klein four-group), all 219 critical points of this truncation exhibit at least $\Z_2\times \Z_2$ symmetry.

Yet another consistent truncation can be found by imposing invariance with respect to all three discrete symmetries in \eqref{Sdef}. The result is a theory with 14 real scalar fields parametrizing the manifold
\be\label{eq:14scman}
\Bigg[\f{\SL(2,\R)}{\U(1)}\Bigg]^7\,.
\ee
The truncation also includes the metric, one gravitino, and seven gaugini. This can be formulated as an $\mathcal{N}=1$ supergravity coupled to 7 chiral multiplets and the potential of this model can be computed analytically in terms of a simple superpotential. This truncation was previously studied in \cite{Guarino:2019jef,Guarino:2020jwv} where 60 critical points were found. Our search has identified 5 new critical points in this truncation all of which  are non-supersymmetric and perturbatively unstable. A very similar truncation in the $\SO(8)$ gauged supergravity was identified and studied in \cite{Bobev:2019dik} and it was shown to contain 48 critical points.

In Table~\ref{tbl:roots} we indicate all root generators that are left invariant in the three truncations summarized above. The seven Cartan generators in \eqref{cartangenerators} are all invariant under $S_1$, $S_2$, and $S_3$.

\section{Numerical search for \texorpdfstring{AdS$_4$}{AdS4} vacua}
\label{sec:vacua}

We now focus on studying the critical points of the potential in the 22-scalar model \eqref{eq:22scman}. The potential for this model can be computed analytically with the help of \texttt{Mathematica} using the solvable parametrization described in Section~\ref{subsec:solvable}. The explicit expression for the potential is unwieldy and will not be presented here. It can be found in the ancillary file \texttt{PotentialAndCriticalPoints.txt} accompanying this arXiv submission. A notable feature of the potential is that it is a function of only 21 of the 22 real scalars in \eqref{eq:22scman}. This is due to an unbroken noncompact symmetry of the potential in this truncation. 

To find the critical points of this potential we resort to numerical techniques. We have employed numerical \texttt{Mathematica} code which uses the explicit analytic form of the potential and its derivatives in combination with the built-in \texttt{FindRoot[$\cdot$]} routine.  After extensive automated searches using this code we have identified 219 distinct critical points.\footnote{A comprehensive \texttt{TensorFlow} search for AdS$_4$ vacua, together with their mass spectra, of the full supergravity potential in \eqref{eq:4dN8potfull}  will be presented in  \cite{BFGP}.}

To present our results, we use the notation \Pu{nnnnnnn} to label a critical point for which the potential evaluates to the numerical value $V=-nn.nnnnn\dots$. Note that these labels are \emph{not} constructed from rounded values of the potential, but rather from the truncated ones. The full list of 219 critical points of the 22 scalar model can be found in Table~\ref{allpoints}. We have also performed a systematic numerical search for critical points of the 14-scalar model in \eqref{eq:14scman}. We have identified 65 critical points which are identified with a $*$ in Table~\ref{allpoints}. We note that the 14-scalar model, and some of its smaller sub-truncations, have been studied before in the literature and as summarized in \cite{Guarino:2020jwv} 60 critical points have been identified. Our list of 65 critical points of the 14-scalar model contains all these 60 critical points.

Some comments are in order. There are two vacua in the full list, both of them previously identified, that have exactly the same value of the potential $V=-2^{16/3}/3^{1/2}$ but are distinct physical solutions. The critical point $\Pu{23277304}_1$ has ${\cal N}=3$ supersymmetry and $\SO(3)\times \SO(3)$ invariance while $\Pu{23277304}_2$ is non-supersymmetric and has G$_2$ invariance. A pair of new AdS$_4$ solutions also deserves special attention. The critical point $\Pu{355983405}$ preserves $\mathcal{N}=1$ supersymmetry and no continuous global symmetry. The critical point $\Pu{355983403}$ is non-supersymmetric, but BF stable, and also does not have a continuous global symmetry. While the first 8 digits in the value of $V$ for these two critical points are identical, we have checked with high precision that they are not the same critical point. The potential for $\Pu{355983405}$  can be computed in a closed algebraic form and is given by $V=-2^{22/3}\times 7^{7/6}/(3\times 5^{5/3})$, while we have not been able to find a similar expression for $\Pu{355983403}$.

For the 219 critical points in Table~\ref{allpoints} we have calculated the mass spectra of all bosonic and fermionic fields of the 4d $\mathcal{N}=8$ supergravity theory. We find that 16 of these AdS$_4$ solutions do not exhibit BF instabilities. We discuss these critical points in more detail in the next section.

\begin{table}[t]
\begin{scriptsize}
\begin{longtable}{lllllllll}
\toprule
$\Pu{19614907}^*$& $\Pu{19987059}^*$& $\Pu{20784609}^*$& $\Pu{21381569}^*$& $\Pu{21867393}^*$& $\Pu{23277304}^*_1$& $\Pu{23277304}^*_2$& $\Pu{23322349}^*$ \\
$\Pu{23413628}^*$& $\Pu{23456052}^*$& $\Pu{23456098}^*$& $\Pu{23456778}^*$& $\Pu{23458779}^*$& $\Pu{23512689}^*$& $\Pu{23715872}$& $\Pu{23795609}^*$ \\
$\Pu{23922493}$& $\Pu{23924535}$& $\Pu{23952870}^*$& $\Pu{24096811}^*$& $\Pu{24149894}$& $\Pu{24149896}^*$& $\Pu{24318391}^*$& $\Pu{24402661}$ \\
$\Pu{24533178}^*$& $\Pu{24691009}^*$& $\Pu{24692967}^*$& $\Pu{24701527}$& $\Pu{25111949}^*$& $\Pu{25693378}^*$& $\Pu{25697101}^*$& $\Pu{25921891}$ \\
$\Pu{25947132}^*$& $\Pu{26884247}$& $\Pu{27101435}^*$& $\Pu{27102980}$& $\Pu{27133412}$& $\Pu{27136806}^*$& $\Pu{27141718}$& $\Pu{27162703}$ \\
$\Pu{27311529}$& $\Pu{27360665}$& $\Pu{27418225}^*$& $\Pu{27450050}$& $\Pu{27609962}^*$& $\Pu{27938983}$& $\Pu{27966612}$& $\Pu{28050399}^*$ \\
$\Pu{28124657}$& $\Pu{28197845}^*$& $\Pu{28210030}$& $\Pu{28486631}^*$& $\Pu{28981494}$& $\Pu{29178187}$& $\Pu{29218859}$& $\Pu{29353962}^*$ \\
$\Pu{29695104}$& $\Pu{29773063}$& $\Pu{29781636}$& $\Pu{29782487}^*$& $\Pu{29984782}$& $\Pu{30083578}$& $\Pu{30116288}$& $\Pu{30241973}^*$ \\
$\Pu{30249480}$& $\Pu{30291641}$& $\Pu{30324163}^*$& $\Pu{30380484}$& $\Pu{30381567}$& $\Pu{30384816}$& $\Pu{30417814}^*$& $\Pu{30434903}^*$ \\
$\Pu{30436799}$& $\Pu{30680917}$& $\Pu{30700546}$& $\Pu{30861770}$& $\Pu{30867190}$& $\Pu{30880584}^*$& $\Pu{31016284}$& $\Pu{31040481}^*$ \\
$\Pu{31195012}^*$& $\Pu{31283092}$& $\Pu{31334051}^*$& $\Pu{31364826}$& $\Pu{31696817}$& $\Pu{31754742}$& $\Pu{31769077}$& $\Pu{31815423}$ \\
$\Pu{31871366}$& $\Pu{31873079}$& $\Pu{32083785}$& $\Pu{32186031}$& $\Pu{32219540}^*$& $\Pu{32253565}$& $\Pu{32254273}$& $\Pu{32390058}$ \\
$\Pu{32416513}$& $\Pu{32628020}$& $\Pu{32634842}$& $\Pu{32784544}$& $\Pu{32816897}^*$& $\Pu{32859039}$& $\Pu{32874514}$& $\Pu{32919810}$ \\
$\Pu{32919845}$& $\Pu{33042513}^*$& $\Pu{33076426}$& $\Pu{33093136}$& $\Pu{33180785}^*$& $\Pu{33181501}^*$& $\Pu{33184214}^*$& $\Pu{33599474}^*$ \\
$\Pu{34064075}$& $\Pu{34215811}$& $\Pu{34478026}$& $\Pu{34757437}^*$& $\Pu{34768017}^*$& $\Pu{34778237}$& $\Pu{34796796}^*$& $\Pu{35001597}^*$ \\
$\Pu{35076002}$& $\Pu{35596894}$& $\Pu{35598271}$& $\Pu{355983403}$& $\Pu{355983405}$& $\Pu{35610181}^*$& $\Pu{35610224}^*$& $\Pu{35610235}^*$ \\
$\Pu{35651026}$& $\Pu{35698097}$& $\Pu{35919737}$& $\Pu{35990424}$& $\Pu{36054725}$& $\Pu{36055521}^*$& $\Pu{36067554}$& $\Pu{36072686}^*$ \\
$\Pu{36092714}$& $\Pu{36100336}$& $\Pu{36144034}$& $\Pu{36228968}$& $\Pu{36234552}$& $\Pu{36329784}$& $\Pu{36944902}$& $\Pu{37258768}$ \\
$\Pu{38153176}$& $\Pu{38539173}^*$& $\Pu{38606422}$& $\Pu{38724006}$& $\Pu{38724625}^*$& $\Pu{38769482}$& $\Pu{38824768}$& $\Pu{38858068}$ \\
$\Pu{39197231}$& $\Pu{39648336}$& $\Pu{39744741}$& $\Pu{39761278}^*$& $\Pu{39765598}$& $\Pu{39794481}$& $\Pu{40592093}$& $\Pu{40600932}^*$ \\
$\Pu{40763269}$& $\Pu{41049034}$& $\Pu{41318714}$& $\Pu{41432947}$& $\Pu{41710023}$& $\Pu{41965792}$& $\Pu{41991457}$& $\Pu{42230669}^*$ \\
$\Pu{42232971}$& $\Pu{42595364}$& $\Pu{43225067}$& $\Pu{43479441}$& $\Pu{43610159}$& $\Pu{44397048}$& $\Pu{44625246}$& $\Pu{44629770}$ \\
$\Pu{44636365}$& $\Pu{44662466}$& $\Pu{47020437}$& $\Pu{47058875}$& $\Pu{47578177}$& $\Pu{48095682}$& $\Pu{48096325}$& $\Pu{48141929}$ \\
$\Pu{48619344}$& $\Pu{49599127}$& $\Pu{49797271}$& $\Pu{50309909}$& $\Pu{52385281}$& $\Pu{52557082}$& $\Pu{53423001}$& $\Pu{53547629}^*$ \\
$\Pu{54934580}$& $\Pu{55416647}$& $\Pu{56156622}$& $\Pu{58641723}$& $\Pu{58768917}$& $\Pu{59496572}$& $\Pu{59571914}$& $\Pu{59603188}$ \\
$\Pu{60664200}$& $\Pu{61221291}$& $\Pu{61898524}^*$& $\Pu{62135276}$& $\Pu{62783730}$& $\Pu{63130245}$& $\Pu{63227985}$& $\Pu{63266652}$ \\
$\Pu{63955920}$& $\Pu{64764264}$& $\Pu{65136976}$& $\Pu{65632203}$& $\Pu{65991426}$& $\Pu{66058194}$& $\Pu{71032966}$& $\Pu{71983709}$ \\
$\Pu{76604976}$& $\Pu{85310001}$& $\Pu{85678094}$&  \\
\bottomrule
\caption{All 219 critical points of the 22-scalar model. The 65 critical points marked with ${}^*$ are also critical points of the 14-scalar model.}
\label{allpoints}
\end{longtable}
\end{scriptsize}
\end{table}

\section{New BF stable \texorpdfstring{AdS$_4$}{AdS4} vacua}
\label{sec:new}

From the list of 16 BF stable critical points we find that only 3 are new. The other 13 perturbatively stable AdS$_4$ solutions lie in the 14-scalar model and have already been given in \cite{Guarino:2020jwv}. Two of the new stable solutions, \Pu{355983403} and \Pu{23715872}, are non-supersymmetric while \Pu{355983405} has ${\cal N}=1$ supersymmetry. In Table~\ref{tableofvacua} we summarize the 16 stable solutions together with the continuous global symmetry and supersymmetry they preserve. We also provide references to the the original literature where the 13 known solutions have been discussed.

In Appendix \ref{sec:sysyAdS4vacua}  we present the full 4d $\mathcal{N}=8$ supergravity mass spectra of the 6 previously known supersymmetric points in Table~\ref{tableofvacua}. In addition we translate this data into information about operator dimension in the dual SCFTs and organize the operators into superconformal multiplets.\footnote{Part of this information was either missing from the previous literature or can be found scattered in different references so we found it useful to collect it in one place.} In Appendix \ref{sec:nonsusy} we similarly give the full four-dimensional mass spectra for all 9 non-supersymmetric in Table~\ref{tableofvacua} points including the 2 new points \Pu{355983403} and \Pu{23715872}. In the ancillary file \texttt{PotentialAndCriticalPoints.txt} attached with this submission we give the mass spectrum for all 219 critical points. We now focus on discussing in more detail the new $\mathcal{N}=1$ AdS$_4$ critical point \Pu{355983405}.

\begin{table}[t]
\begin{center}
\begin{tabular}{@{\extracolsep{10 pt}}l l l c c}
\toprule
 Point& SUSY & Cont. symmetry&  $V$ & Reference \\
\midrule
\noalign{\smallskip}
\hyperref[P19987059]{\Pu{19987059}} 		& $\mathcal{N}=1$ & ${\rm G}_2$ &  $-\frac{2^{28/3} \times 3^{1/2}}{5^{5/2}}$& \cite{Behrndt:2004km,Borghese:2012qm}\\
\hyperref[P20784609]{\Pu{20784609}} 		& $\mathcal{N}=2$ & $\SU(3) \times \U(1)$ & $-2^2 \times 3^{3/2}$ &\cite{Guarino:2015jca}\\
\hyperref[P23277304-1]{\Pu{23277304}$_1$} 	& $\mathcal{N}=3$ & $\SO(3)\times \SO(3)$ &  $-\tfrac{2^{16/3}}{3^{1/2}}$ &\cite{Gallerati:2014xra}\\
\hyperref[P23795609]{\Pu{23795609}} 		& $\mathcal{N}=1$ & $\SU(3)$ & $-\tfrac{2^8\times 3^{3/2}}{5^{5/2}}$ & \cite{Guarino:2015qaa}\\
\hyperref[P25697101]{\Pu{25697101}} 		& $\mathcal{N}=1$ & $\U(1)$ & $-25.697101$ & \cite{Guarino:2019snw}\\
\hyperref[P355983405]{\Pu{355983405}}	 	& $\mathcal{N}=1$ & $\emptyset$ & $-\tfrac{2^{22/3}\times 7^{7/6}}{3\times 5^{5/3}}$ & Here\\
\hyperref[P35610235]{\Pu{35610235}} 		& $\mathcal{N}=1$ & $\U(1)$ & $-35.610235$ & \cite{Guarino:2019snw}\\
\midrule
\hyperref[P23277304-2]{\Pu{23277304}$_2$} 	& $\mathcal{N}=0$ &	${\rm G}_2$ 			&	$-\tfrac{2^{16/3}}{3^{1/2}}$ & \cite{Lust:2008zd,Borghese:2012qm}\\
\hyperref[P23413628]{\Pu{23413628}}			& $\mathcal{N}=0$ &	$\SU(3)$ 				&	$-23.413628$ & \cite{Guarino:2015qaa}\\
\hyperref[P23456052]{\Pu{23456052}}			& $\mathcal{N}=0$ & $\SO(3)\times \U(1)$	&	$-23.456053$ & \cite{Guarino:2019jef}\\
\hyperref[P23456098]{\Pu{23456098}}			& $\mathcal{N}=0$ & $\SO(3)$				&	$-23.456098$ & \cite{Guarino:2019jef}\\
\hyperref[P23456778]{\Pu{23456778}}			& $\mathcal{N}=0$ & $\SU(3)$				&	$-23.456779$ & \cite{Guarino:2015qaa}\\
\hyperref[P23458779]{\Pu{23458779}}			& $\mathcal{N}=0$ & $\SO(3)\times \U(1)$	&	$-23.458780$ & \cite{Guarino:2019jef}\\
\hyperref[P23512689]{\Pu{23512689}}			& $\mathcal{N}=0$ & $\SO(3)\times \SO(3)$	&	$-23.512690$ & \cite{Guarino:2015qaa}\\
\hyperref[P23715872]{\Pu{23715872}}			& $\mathcal{N}=0$ & $\U(1)$					&	$-23.715872$ & Here\\
\hyperref[P355983403]{\Pu{355983403}}		& $\mathcal{N}=0$ & $\emptyset$				&	$-35.5983403$ & Here\\
\bottomrule
\end{tabular}
\caption{All known BF stable AdS$_4$ solutions in the 22-scalar truncation \eqref{eq:22scman} including the new solutions \Pu{355983403}, \Pu{355983405}, and \Pu{23715872}.}
\label{tableofvacua}
\end{center}
\end{table}

\subsection{Supergravity mass spectra for \Pu{355983405}}
\label{P355983405}

The new $\mathcal{N}=1$ AdS$_4$ solution \Pu{355983405} has the following algebraic value of the potential
\begin{equation}
V = -\frac{2^{22/3}\times 7^{7/6}}{3\times 5^{5/3}}\,.
\end{equation}
We arrived at this algebraic expression by comparing it to the numerical value determined by our numerical codes with an accuracy of more than 300 digits. The constant values of the scalars fields which specify the location of the critical point on the scalar manifold can be found in the ancillary file \texttt{PotentialAndCriticalPoints.txt} attached with this submission.  It should be noted that in the solvable parametrization of the scalar coset described above the location of the critical point is determined by 17 non-trivial values for the 22 scalar fields in \eqref{eq:22scman}. To determine some of the properties of this critical point it is instructive to calculate the masses for the linearized perturbation of bosonic and fermionic fields in the 4d $\mathcal{N}=8$ supergravity theory. To this end we have used the mass formulae summarized in \cite{Gallerati:2014xra} and computed the mass spectra numerically to a high degree of accuracy. For clarity, below we present only the first 7 digits for each mass. We use the subscripts $\psi$, $A$, $\chi$, and $\phi$ to denote the 8 gravitini, 28 spin-1 fields, 56 spin-1/2 fields, and the 70 scalars in the 4d $\mathcal{N}=8$ supergravity, respectively. To indicate the degeneracy, $n$, of each of the modes we use a subscript $_{\times n}$ next to the numerical value of the mass. We then find the following values for the dimensionless squared masses $m^2L^2$ for the \Pu{355983405} critical point\footnote{We use conventions in which the AdS$_4$ length scale, $L$, is related to the potential as $V = -6/L^2$.}
\begin{flalign*}
\qquad\qquad m^2_\psi L^2 &: 5.02803_{\times 1}\,,~4.89758_{\times 1}\,,~3.25582_{\times 1}\,,~3.14936_{\times 1}\,,~2.75003_{\times 1}\,,~ &\\
&\quad 2.57143_{\times 1}\,,~2.41918_{\times 1}\,,~1_{\times 1}\,,~ &\\
m^2_A L^2 &: 7.27036_{\times 1}\,,~7.19910_{\times 1}\,,~7.11063_{\times 1}\,,~6.95606_{\times 1}\,,~6.35686_{\times 1}\,,~ &\\
&\quad 5.97573_{\times 1}\,,~5.78845_{\times 1}\,,~5.67796_{\times 1}\,,~5.21669_{\times 1}\,,~5.06021_{\times 1}\,,~ &\\
&\quad 4.92400_{\times 1}\,,~4.73855_{\times 1}\,,~4.40835_{\times 1}\,,~4.27033_{\times 1}\,,~4.17500_{\times 1}\,,~ &\\
&\quad 3.97455_{\times 1}\,,~2.78570_{\times 1}\,,~2.68453_{\times 1}\,,~1.45143_{\times 1}\,,~1.37472_{\times 1}\,,~ &\\
&\quad 1.09171_{\times 1}\,,~0.967861_{\times 1}\,,~0.863807_{\times 1}\,,~0.672028_{\times 1}\,,~0.668804_{\times 1}\,,~ &\\
&\quad 0.394356_{\times 1}\,,~0.367529_{\times 1}\,,~0.00325020_{\times 1}\,,~ &\\
m^2_\chi L^2 &: 20.1121_{\times 1}\,,~19.5903_{\times 1}\,,~13.0233_{\times 1}\,,~12.5974_{\times 1}\,,~11.0001_{\times 1}\,,~ &\\
&\quad 10.4284_{\times 1}\,,~10.2857_{\times 1}\,,~10.1405_{\times 1}\,,~9.87480_{\times 1}\,,~9.67671_{\times 1}\,,~ &\\
&\quad 9.42724_{\times 1}\,,~9.42510_{\times 1}\,,~8.97087_{\times 1}\,,~8.74578_{\times 1}\,,~8.74207_{\times 1}\,,~ &\\
&\quad 8.61270_{\times 1}\,,~8.05479_{\times 1}\,,~7.62328_{\times 1}\,,~7.47206_{\times 1}\,,~7.36171_{\times 1}\,,~ &\\
&\quad 7.01399_{\times 1}\,,~6.89643_{\times 1}\,,~5.02803_{\times 1}\,,~4.96980_{\times 1}\,,~4.89758_{\times 1}\,,~ &\\
&\quad 4.77165_{\times 1}\,,~4.28648_{\times 1}\,,~4.24572_{\times 1}\,,~4.22499_{\times 1}\,,~4.00432_{\times 1}\,,~ &\\
&\quad 3.98059_{\times 1}\,,~3.83113_{\times 1}\,,~3.74322_{\times 1}\,,~3.37860_{\times 1}\,,~3.25582_{\times 1}\,,~ &\\
&\quad 3.14936_{\times 1}\,,~3.00505_{\times 1}\,,~2.75003_{\times 1}\,,~2.64422_{\times 1}\,,~2.57143_{\times 1}\,,~ &\\
&\quad 2.41918_{\times 1}\,,~2.13225_{\times 1}\,,~2.12735_{\times 1}\,,~1.69707_{\times 1}\,,~1.65336_{\times 1}\,,~ &\\
&\quad 1.00649_{\times 1}\,,~0.866994_{\times 1}\,,~0.283967_{\times 1}\,,~0.275774_{\times 1}\,,~0.274515_{\times 1}\,,~ &\\
&\quad 0.211805_{\times 1}\,,~0.210261_{\times 1}\,,~0.139918_{\times 1}\,,~0.0916383_{\times 1}\,,~0.0816989_{\times 1}\,,~ &\\
&\quad 0.0000105000_{\times 1}\,,~ &\\
m^2_\phi L^2 &: 11.0172_{\times 1}\,,~10.4951_{\times 1}\,,~9.69877_{\times 1}\,,~8.38431_{\times 1}\,,~8.07495_{\times 1}\,,~ &\\
&\quad 7.66239_{\times 1}\,,~5.19910_{\times 1}\,,~4.95606_{\times 1}\,,~4.73238_{\times 1}\,,~4.35686_{\times 1}\,,~ &\\
&\quad 4.35507_{\times 1}\,,~4.30623_{\times 1}\,,~4.28046_{\times 1}\,,~4.00540_{\times 1}\,,~3.97573_{\times 1}\,,~ &\\
&\quad 3.78845_{\times 1}\,,~3.78537_{\times 1}\,,~3.67796_{\times 1}\,,~3.21669_{\times 1}\,,~2.86225_{\times 1}\,,~ &\\
&\quad 2.73855_{\times 1}\,,~2.64846_{\times 1}\,,~2.36560_{\times 1}\,,~2.27033_{\times 1}\,,~0.185203_{\times 1}\,,~ &\\
&\quad 0.169510_{\times 1}\,,~0.00324116_{\times 1}\,,~0_{\times 28}\,,~-0.201881_{\times 1}\,,~-1.18315_{\times 1}\,,~ &\\
&\quad -1.19908_{\times 1}\,,~-1.20154_{\times 1}\,,~-1.32797_{\times 1}\,,~-1.33120_{\times 1}\,,~-1.48603_{\times 1}\,,~ &\\
&\quad -1.60564_{\times 1}\,,~-1.63247_{\times 1}\,,~-1.99675_{\times 1}\,,~-2.06413_{\times 1}\,,~-2.23414_{\times 1}\,,~ &\\
&\quad -2.24892_{\times 1}\,,~-2.24937_{\times 1}\,,~-2.24943_{\times 1}\,,~ &\end{flalign*}
From the gravitino mass spectrum we read of that there is one spin-3/2 mode of mass $m^2_\psi L^2=1$ corresponding to the preserved $\mathcal{N}=1$ supersymmetry. The critical point is invariant under the $\Z_2\times \Z_2$ symmetry used to specify the truncation in \eqref{eq:22scman}. There is no continuous symmetry since there are no massless spin-1 modes in the spectrum. This implies that the  $\ISO(7)$ gauge symmetry of the 4d $\mathcal{N}=8$ supergravity is completely broken. This is further supported by the fact that there are 28 massless Goldstone scalars.

\subsection{3d \texorpdfstring{${\cal N}=1$}{N=1} CFT spectrum for \Pu{355983405}}

The \Pu{355983405} AdS$_4$ solution can be consistently embedded in type IIA string theory and thus should have a well-defined 3d $\mathcal{N}=1$ SCFT as a holographic dual. This should be a strongly coupled CFT with no continuous global symmetry and a minimal amount of supersymmetry, which  makes it hard to study with conventional QFT techniques. Therefore it is valuable to use the AdS/CFT dictionary and the mass spectra computed above to calculate the spectrum of low-dimensional operators in this theory. The formulae relating supergravity masses to conformal dimensions for operators of different spin are summarized in Table~\ref{tbl-dims}. The spectrum of operator dimensions can then be organized into multiplets of the 3d $\mathcal{N}=1$ superconformal algebra. 

Before we present the results of this calculation we comment on some general features.\footnote{For the 6 other supersymmetric critical points in Table~\ref{tableofvacua} the SCFT operator spectra and their organization into superconformal multiplets is presented in Appendix~\ref{sec:sysyAdS4vacua}.} Some of the modes in the full spectrum of quadratic fluctuations do not correspond to CFT operators since they are ``eaten'' by the usual (super-)Higgs mechanism. In particular, if the $\ISO(7)$ gauge symmetry of the 4d $\mathcal{N}=8$ supergravity is broken to a subgroup $\mathfrak{g}$ in a given AdS$_4$ vacuum, then the number of massless spin-0 modes that combine with the massive vectors is $28-\dim\mathfrak{g}$. Similarly, for each massive gravitino with mass $m_\psi^2L^2 > 1$ a spin-$1/2$ fermion with mass $m_\chi^2 = 4m_\psi^2$ is eaten. For the solution \Pu{355983405} there is no continuous symmetry left and thus 28 massless scalars and seven massive spin-1/2 fermions are taken by the Higgs mechanism. Only after removing these Goldstone modes from the mass spectrum one should apply the formulae in Table \ref{tbl-dims} to determine the SCFT operator spectrum.
\begin{table}[t]
\begin{center}
\begin{tabular}{@{\extracolsep{15 pt}} l l }
\toprule
\quad Spin & Dimension    \\
\midrule
\noalign{\smallskip}
\quad0 & $\Delta =\frac{3}{2}\pm \sqrt{\frac{9}{4}+m^2L^2}$ \\
\quad$\frac{1}{2}$ & $\Delta  =\frac{3}{2}+|mL|$ \\
\quad$1$ & $\Delta =\frac{3}{2}\pm \sqrt{\frac{1}{4}+m^2L^2}$ \\
\quad$\frac{3}{2}$ & $\Delta =\frac{3}{2}+|mL|$ \\
\bottomrule
\end{tabular}
\caption{Conformal dimensions of CFT operators dual to supergravity fields of spin, $s$, and mass, $m$. }
\label{tbl-dims}
\end{center}
\end{table}
Furthermore, when computing the dimensions of operators dual to  spin-1 and spin-0 modes we must be careful to choose the appropriate signs in the formulae in Table \ref{tbl-dims} such that the unitarity bound is obeyed. For some of the scalar modes two possible dimensions are compatible with the unitarity bound and one has a choice of alternate quantization, see \cite{Klebanov:1999tb}. This choice is unambiguously fixed by organizing the operator spectrum into ${\cal N}=1$ CFT multiplets.

There are two types of long and one type of short 3d ${\cal N}=1$ superconformal multiplets we encounter. They will be denoted as follows\footnote{These multiplets are denoted as $A_1$, $L'$, and $L$ in \cite{Cordova:2016emh}, respectively. The multiplets $A_2'$ and $B_1$ in \cite{Cordova:2016emh} correspond to a free $\mathcal{N}=1$ chiral field and the identity operator, respectively, and will not play a role here.}
\be
\begin{split}\label{eq:3dN1multiplets}
\text{Short}[s>0] &= \big\{|s+1,s\rangle,|s+\tfrac32,s+\tfrac12\rangle \big\}\,,\\
\text{Long}[\Delta,0] &= \big\{|\Delta,0\rangle,|\Delta+\tfrac12,\tfrac12\rangle,|\Delta+1,0\rangle \big\}\,,\\
\text{Long}[\Delta,s>0] &= \big\{|\Delta,s\rangle,|\Delta+\tfrac12,s+\tfrac12\rangle,|\Delta+\tfrac12,s-\tfrac12\rangle,|\Delta+1,s\rangle \big\}\,,
\end{split}
\ee
where we use $|\Delta,s\rangle$ to denote a CFT operator with conformal dimension $\Delta$ and spin $s$. Note that the long multiplets are constrained by the unitarity bound $\Delta>s+1$. All 3d SCFTs contain a single short multiplet consisting of the energy-momentum tensor and the supercurrent
\be
\text{Short}[\tfrac32]\,.
\ee
The remaining superconformal multiplets can be determined algorithmically by organizing the operator dimensions according to \eqref{eq:3dN1multiplets}. Carrying out this procedure, we arrive at the following list
\be
\begin{split}
&\text{Long}[3.24233,1]\,,\qquad \text{Long}[3.21305,1]\,,\qquad \text{Long}[2.80439,1]\,,\\
&\text{Long}[2.77464,1]\,,\qquad \text{Long}[2.65832,1]\,,\qquad \text{Long}[2.60357,1]\,,\\
&\text{Long}[2.55537,1]\,,\qquad \text{Long}[3.72930,\tfrac12]\,,\qquad \text{Long}[3.68441,\tfrac12]\,,\\
&\text{Long}[3.57038,\tfrac12]\,,\qquad \text{Long}[3.49514,\tfrac12]\,,\qquad \text{Long}[3.45733,\tfrac12]\,,\\
&\text{Long}[3.43474,\tfrac12]\,,\qquad \text{Long}[3.33810,\tfrac12]\,,\qquad \text{Long}[3.23351,\tfrac12]\,,\\
&\text{Long}[3.12611,\tfrac12]\,,\qquad \text{Long}[1.96022,\tfrac12]\,,\qquad \text{Long}[1.95854,\tfrac12]\,,\\
&\text{Long}[1.80272,\tfrac12]\,,\qquad \text{Long}[1.78583,\tfrac12]\,,\qquad \text{Long}[1.50324,\tfrac12]\,,\\
&\text{Long}[4.14242,0]\,,\qquad \text{Long}[4.07003,0]\,,\qquad \text{Long}[3.95670,0]\,,\\
&\text{Long}[3.76103,0]\,,\qquad \text{Long}[3.71325,0]\,,\qquad \text{Long}[3.64839,0]\,,\\
&\text{Long}[3.06051,0]\,,\qquad \text{Long}[3.05548,0]\,,\qquad \text{Long}[3.00108,0]\,,\\
&\text{Long}[1.93113,0]\,,\qquad \text{Long}[1.53289,0]\,,\qquad \text{Long}[1.52514,0]\,,\\
&\text{Long}[1.52394,0]\,,\qquad \text{Long}[1.37406,0]\,.
\end{split}
\ee
Notice that all of these multiplets are long, and therefore unprotected, which is compatible with the expected behavior of a minimally supersymmetric SCFT with no continuous global symmetry.

\bigskip
\bigskip
\leftline{\bf Acknowledgements}
\smallskip
\noindent 
We  would like to thank Jesse van Muiden for useful discussions. The work of NB is supported in part by an Odysseus grant G0F9516N from the FWO and by the KU Leuven C1 grant ZKD1118 C16/16/005. FFG is supported by the University of Iceland Recruitment Fund and the Research Foundation - Flanders, FWO. KP is supported in part by DOE grant DE-SC0011687. 



\appendix
\section{Root generators of \texorpdfstring{$\Ess$}{E7(7)}}
\label{sec:roots}

\begin{footnotesize}
\begin{longtable}{@{\extracolsep{20 pt}}lll ccc}
\toprule
$\alpha\in\Delta_+$ & $\mathfrak{e}_\alpha$ & $\mathfrak{f}_\alpha$  & $S_1$ & $S_1S_2$ & $S_1S_2S_3$ \\
\midrule
\noalign{\smallskip}\endhead
\text{[1000000]} & $t{}_1{}^2$ & $t{}_2{}^1$ &$*$&&\\
\text{[0100000]} & $t{}_2{}^3$ & $t{}_3{}^2$ &$*$&$*$& \\
\text{[0010000]} & $t{}_3{}^4$ & $t{}_4{}^3$ &&&\\
\text{[0001000]} & $t{}_4{}^5$ & $t{}_5{}^4$ &$*$&$*$&\\
\text{[0000100]} & $t{}_5{}^6$ & $t{}_6{}^5$ &$*$&&\\
\text{[0000010]} & $t{}_6{}^7$ & $t{}_7{}^6$ &$*$&$*$&\\
\text{[0000001]} & $t_{5678}$ & $t_{1234}$   &&&
\\[3pt]
\text{[1100000]} & $t{}_1{}^3$ & $t{}_3{}^1$ &$*$&&\\
\text{[0110000]} & $t{}_2{}^4$ & $t{}_4{}^2$ &&&\\
\text{[0011000]} & $t{}_3{}^5$ & $t{}_5{}^3$ &&&\\
\text{[0001100]} & $t{}_4{}^6$ & $t{}_6{}^4$ &$*$&&\\
\text{[0000110]} & $t{}_5{}^7$ & $t{}_7{}^5$ &$*$&&\\
\text{[0001001]} & $t_{4678}$ & $t_{1235}$   &&&
\\[3pt]
\text{[1110000]} & $t{}_1{}^4$ & $t{}_4{}^1$ &&&\\
\text{[0111000]} & $t{}_2{}^5$ & $t{}_5{}^2$ &&&\\
\text{[0011100]} & $t{}_3{}^6$ & $t{}_6{}^3$ &&&\\
\text{[0001110]} & $t{}_4{}^7$ & $t{}_7{}^4$ &$*$&&\\
\text{[0001101]} & $t_{4578}$ & $t_{1236}$   &&&\\
\text{[0011001]} & $t_{3678}$ & $t_{1245}$   &$*$&&
\\[3pt]
\text{[1111000]} & $t{}_1{}^5$ & $t{}_5{}^1$ &&&\\
\text{[0111100]} & $t{}_2{}^6$ & $t{}_6{}^2$ &&&\\
\text{[0011110]} & $t{}_3{}^7$ & $t{}_7{}^3$ &&&\\
\text{[0001111]} & $t_{4568}$ & $t_{1237}$   &&&\\
\text{[0011101]} & $t_{3578}$ & $t_{1246}$   &$*$&$*$&\\
\text{[0111001]} & $t_{2678}$ & $t_{1345}$   &$*$&&
\\[3pt]
\text{[1111100]} & $t{}_1{}^6$ & $t{}_6{}^1$ &&&\\
\text{[0111110]} & $t{}_2{}^7$ & $t{}_7{}^2$ &&&\\
\text{[0011111]} & $t_{3567}$ & $t_{1247}$   &$*$&$*$&$*$\\
\text{[0012101]} & $t_{3478}$ & $t_{1256}$   &$*$&$*$&$*$\\
\text{[0111101]} & $t_{2578}$ & $t_{1346}$   &$*$&$*$&$*$\\
\text{[1111001]} & $t_{1678}$ & $t_{2345}$   &$*$&$*$&$*$ 
\\[3pt]
\text{[1111110]} & $t{}_1{}^7$ & $t{}_7{}^1$ &&&\\
\text{[0012111]} & $t_{3468}$ & $t_{1257}$   &$*$&$*$&\\
\text{[0111111]} & $t_{2568}$ & $t_{1347}$   &$*$&$*$&\\
\text{[0112101]} & $t_{2478}$ & $t_{1356}$   &$*$&$*$&\\
\text{[1111101]} & $t_{1578}$ & $t_{2346}$   &$*$&&
\\[3pt]
\text{[0012211]} & $t_{3458}$ & $t_{1267}$   &$*$&&\\
\text{[0112111]} & $t_{2468}$ & $t_{1357}$   &$*$&$*$&$*$\\
\text{[0122101]} & $t_{2378}$ & $t_{1456}$   &&&\\
\text{[1112101]} & $t_{1478}$ & $t_{2356}$   &$*$&&\\
\text{[1111111]} & $t_{1568}$ & $t_{2347}$   &$*$&&
\\[3pt]
\text{[0112211]} & $t_{2458}$ & $t_{1367}$   &$*$&&\\
\text{[1122101]} & $t_{1378}$ & $t_{2456}$   &&&\\
\text{[0122111]} & $t_{2368}$ & $t_{1457}$   &&&\\
\text{[1112111]} & $t_{1468}$ & $t_{2357}$   &$*$&&
\\[3pt]
\text{[1222101]} & $t_{1278}$ & $t_{3456}$   &&&\\
\text{[1122111]} & $t_{1368}$ & $t_{2457}$   &&&\\
\text{[1112211]} & $t_{1458}$ & $t_{2367}$   &$*$&$*$&$*$\\
\text{[0122211]} & $t_{2358}$ & $t_{1467}$   &&&
\\[3pt]
\text{[1222111]} & $t_{1268}$ & $t_{3457}$   &&&\\
\text{[1122211]} & $t_{1358}$ & $t_{2467}$   &&&\\
\text{[0123211]} & $t_{2348}$ & $t_{1567}$   &&&
\\[3pt]
\text{[0123212]} & $t{}_8{}^1$ & $t{}_1{}^8$ &$*$&$*$&\\
\text{[1222211]} & $t_{1258}$ & $t_{3467}$   &&&\\
\text{[1123211]} & $t_{1348}$ & $t_{2567}$   &&&
\\[3pt]
\text{[1123212]} & $t{}_8{}^2$ & $t{}_2{}^8$ &$*$&&\\
\text{[1223211]} & $t_{1248}$ & $t_{3567}$   &&&
\\[3pt]
\text{[1223212]} & $t{}_8{}^3$ & $t{}_3{}^8$ &$*$&& \\
\text{[1233211]} & $t_{1238}$ & $t_{4567}$   &$*$&$*$&$*$
\\[3pt]
\text{[1233212]} & $t{}_8{}^4$ & $t{}_4{}^8$ &&&
\\[3pt]
\text{[1234212]} & $t{}_8{}^5$ & $t{}_5{}^8$ &&&
\\[3pt]
\text{[1234312]} & $t{}_8{}^6$ & $t{}_6{}^8$ &&&
\\[3pt]
\text{[1234322]} & $t{}_8{}^7$ & $t{}_7{}^8$ &&&\\
\bottomrule
\caption{\label{tbl:roots} Positive and negative root generators of $\Ess$ with respect to the Cartan subalgebra \eqref{cartangenerators}. In the last three columns we indicate with a $*$ which root generators are invariant under the discrete symmetry actions $S_1$, $S_1S_2$, and $S_1S_2S_3$.}\\
\end{longtable}
\end{footnotesize}

\section{The spectra of supersymmetric \texorpdfstring{AdS$_4$}{AdS4} vacua}
\label{sec:sysyAdS4vacua}

Here we present the spectra of masses around the 6 supersymmetric AdS$_4$ vacua in Table~\ref{tableofvacua} and arrange them in superconformal multiplets. The spectrum of the new $\mathcal{N}=1$ vacuum \Pu{355983405} is given in Section~\ref{sec:new}.

\paragraph{\Pu{19987059}} This is an ${\cal N}=1$ vacuum with
\label{P19987059}
\be
V=-\frac{2^{28/3}\times 3^{1/2}}{5^{5/2}}\,,
\ee
and ${\rm G}_2$ continuous symmetry first discovered in \cite{Behrndt:2004km} in massive type IIA supergravity. It was later rediscovered in four-dimensional supergravity in \cite{Borghese:2012qm} where the bosonic spectrum in  four dimensions was determined. The complete four-dimensional mass spectrum is
\begin{flalign*}
\qquad\qquad m^2_\psi L^2 &: ({\tfrac32})_{\times 7}\,,~ 1_{\times 1}\\
m^2_A L^2 &: (\tfrac32+\sqrt{\tfrac32})_{\times 7}\,,~ (\tfrac32-\sqrt{\tfrac32})_{\times 7}\,,~ 0_{\times 14}\,,&\\
m^2_\chi L^2 &: 6_{\times 8}\,,~(\tfrac32)_{\times 7}\,,~(\tfrac16)_{\times 27}\,,~0_{\times 14}\,,&\\
m^2_\phi L^2 &: (4+\sqrt{6})_{\times 1}\,,~ (4-\sqrt{6})_{\times 1},~ 0_{\times 14}\,,~ (-\tfrac{11}6+\sqrt{\tfrac16})_{\times 27}\,,~ (-\tfrac{11}6-\sqrt{\tfrac16})_{\times 27}\,.&
\end{flalign*} 
We can translate this into conformal dimensions for the dual 3d CFT operators and organize them in superconformal multiplets as described in Section \ref{sec:new}. This results in 2 short
\be
\text{Short}[\tfrac32]_{\times 1}\,,\quad \text{Short}[\tfrac12]_{\times 14}\,,
\ee
and 3 long multiplets
\be
\text{Long}\Big[1+\sqrt{\tfrac32},1\Big]_{\times 7}\,,\quad \text{Long}\Big[1+\sqrt{6},0\Big]_{\times 1}\,,\quad \text{Long}\Big[1+\sqrt{\tfrac16},0\Big]_{\times 27}\,.
\ee

\paragraph{\Pu{20784609}} This is an ${\cal N}=2$ vacuum with
\label{P20784609}
\be
V=-2^2\times 3^{3/2}\,,
\ee
and $\SU(3)\times\U(1)$ continuous symmetry. The solution was first studied in \cite{Guarino:2015jca} and the four-dimensional bosonic spectrum was computed in \cite{Guarino:2015qaa}. The complete four-dimensional mass spectrum including also fermionic fluctuations is
\begin{flalign*}
\qquad\qquad m^2_\psi L^2 &: (\tfrac{16}9)_{\times 6}\,,~ 1_{\times 2}\\
m^2_A L^2 &:  4_{\times 1}\,,~(\tfrac{28}9)_{\times 6}\,,~ (\tfrac49)_{\times 12}\,,~ 0_{\times 9}\,,&\\
m^2_\chi L^2 &: (\tfrac{64}{9})_{\times 6}\,,~(\tfrac92 + \sqrt{\tfrac{17}4})_{\times 2}\,,~(\tfrac92 - \sqrt{\tfrac{17}4})_{\times 2}\,,~(\tfrac{16}9)_{\times 12}\,,~(\tfrac19)_{\times 18}\,,~0_{\times 16}\,,&\\
m^2_\phi L^2 &: (3+\sqrt{17})_{\times 1}\,,~ 2_{\times 3},~ 0_{\times 19}\,,~ (3-\sqrt{17})_{\times 1}\,,~ (-\tfrac{14}9)_{\times 18}\,,~ (-2)_{\times 16}\,,~ (-\tfrac{20}9)_{\times 12}\,.
\end{flalign*}
To map this supergravity mass spectrum to 3d $\mathcal{N}=2$ superconformal multiplets we use the results and notation in \cite{Cordova:2016emh}. The SCFT at hand has the following spectrum of low lying operators\footnote{We use the notation $|\Delta,s\rangle$ to denote individual operators in a given multiplet.}

\begin{itemize}

\item The $\cN=2$ energy-momentum tensor multiplet $A_1\bar{A}_1[2]_2^{(0)}$
\begin{equation}
|2,1\rangle \,, \qquad |\tfrac{5}{2},\tfrac{3}{2}\rangle \times 2\,, \qquad |3,2\rangle\,.
\end{equation}

\item 8 $A_2\bar{A}_2[0]_1^{(0)}$ conserved current multiplets corresponding to the  $\SU(3)$ flavor symmetry
\begin{equation}
|1,0\rangle \,, \qquad |\tfrac{3}{2},\tfrac{1}{2}\rangle \times 2\,, \qquad |2,1\rangle \,, \qquad |2,0\rangle\,.
\end{equation}

\item 6  $L\bar{A}[1]_{\frac{11}{6}}$ semi-short multiplets
\begin{equation}
\begin{split}
|\tfrac{11}{6},\tfrac{1}{2}\rangle \,, \qquad |\tfrac{7}{3},1\rangle\times 2 \,, \qquad |\tfrac{7}{3},0\rangle \,, \qquad |\tfrac{17}{6},\tfrac{1}{2}\rangle\times 2 \,, \qquad |\tfrac{17}{6},\tfrac{3}{2}\rangle \,, \qquad |\tfrac{10}{3},1\rangle \,.
\end{split}
\end{equation}

\item 12  $L\bar{B}[0]_{\frac{4}{3}}$ semi-short multiplets
\begin{equation}
\begin{split}
|\tfrac{4}{3},0\rangle \,, \qquad |\tfrac{11}{6},\tfrac{1}{2}\rangle \,, \qquad  |\tfrac{7}{3},0\rangle \,.
\end{split}
\end{equation}

\item 1 long  $L\bar{L}[0]_{\frac{1}{2}+\frac{\sqrt{17}}{2}}$ multiplet 
\begin{equation}
\begin{split}
&|\tfrac{1}{2}+\tfrac{\sqrt{17}}{2},0\rangle \,, \qquad |1+\tfrac{\sqrt{17}}{2},\tfrac{1}{2}\rangle\times 2 \,, \qquad |\tfrac{3}{2}+\tfrac{\sqrt{17}}{2},0\rangle\times 3 \,, \\
&|\tfrac{3}{2}+\tfrac{\sqrt{17}}{2},1\rangle \,, \qquad |2+\tfrac{\sqrt{17}}{2},\tfrac{1}{2}\rangle \times 2 \,, \qquad |\tfrac{5}{2}+\tfrac{\sqrt{17}}{2},0\rangle \,.
\end{split}
\end{equation}
\end{itemize}
The full KK spectrum of massive IIA supergravity for this AdS$_4$ vacuum was recently computed in \cite{Varela:2020wty} using the techniques developed in \cite{Malek:2019eaz,Malek:2020yue}.

\paragraph{\Pu{23277304}$_1$} This is an ${\cal N}=3$ vacuum with
\label{P23277304-1}
\be
V=-\frac{2^{16/3}}{3^{1/2}}\,,
\ee
and $\SU(2)\times\SU(2)$ continuous symmetry. The solution was first studied in \cite{Gallerati:2014xra} where the four-dimensional spectrum was computed. We repeat it here
\begin{flalign*}
\qquad\qquad m^2_\psi L^2 &: 3_{\times 1}\,,~ (\tfrac94)_{\times 4}\,,~ 1_{\times 3}\\
m^2_A L^2 &:  (3+\sqrt{3})_{\times 3}\,,~(\tfrac{15}4)_{\times 4}\,,~ (3-\sqrt{3})_{\times 3}\,,~(\tfrac34)_{\times 12}\,,~ 0_{\times 6}\,,&\\
m^2_\chi L^2 &: 12_{\times 1}\,,~9_{\times 4}\,,~(4+2\sqrt{3})_{\times 3}\,,~3_{\times 8}\,,~(\tfrac94)_{\times 12}\,,~(4-2\sqrt{3})_{\times 3}\,, ~(\tfrac14)_{\times 12}\,,~0_{\times 13}\,,&\\
m^2_\phi L^2 &: (3+3\sqrt{3})_{\times 1}\,,~ (1+\sqrt{3})_{\times 6},~ 0_{\times 22}\,,~ (1-\sqrt{3})_{\times 6}\,,&\\
&\quad (-\tfrac{5}4)_{\times 12}\,,~ (-2)_{\times 18}\,,~ (3-3\sqrt{3})_{\times 1}\,,~ (-\tfrac94)_{\times 4}\,.
\end{flalign*}
To map the supergravity mass spectrum to 3d $\mathcal{N}=3$ superconformal multiplets we use the results and notation in \cite{Cordova:2016emh}. The SCFT at hand has the following low-lying spectrum

\begin{itemize}

\item The $\cN=3$ EM tensor multiplet $A_1[1]_{\frac{3}{2}}^{(0)}$
\begin{equation}
|\tfrac{3}{2},\tfrac{1}{2}\rangle\,, \qquad |2,1\rangle\times 3 \,, \qquad |\tfrac{5}{2},\tfrac{3}{2}\rangle \times 3\,, \qquad |3,2\rangle\,.
\end{equation}

\item 3 $B_1[0]_1^{(2)}$ current multiplets corresponding to the $\SO(3)$ flavor symmetry
\begin{equation}
|1,0\rangle\times 3 \,, \qquad |\tfrac{3}{2},\tfrac{1}{2}\rangle \times 4\,, \qquad |2,1\rangle \,, \qquad |2,0\rangle\times 3\,.
\end{equation}

\item 2 $A_2[0]_{\frac{3}{2}}^{(1)}$ semi-short multiplets
\begin{equation}
\begin{split}
 &|\tfrac{3}{2},0\rangle \times 2\,, \qquad |2,\tfrac{1}{2}\rangle\times 6 \,, \qquad |\tfrac{5}{2},1\rangle\times 6\,, \qquad |\tfrac{5}{2},0\rangle\times 6\,, \\
 & |3,\tfrac{3}{2}\rangle\times 2\,, \qquad |3,\tfrac{1}{2}\rangle\times 6\,, \qquad |\tfrac{7}{2},1\rangle\times 2\,.
\end{split}
\end{equation}

\item 1 long $L[0]_{\sqrt{3}}^{(0)}$ multiplet with the following content
\begin{equation}
\begin{split}
 &|\sqrt{3},0\rangle\,, \quad |\tfrac{1}{2}+\sqrt{3},\tfrac{1}{2}\rangle \times 3\,, \qquad |1+\sqrt{3},0\rangle \times 6 \, \quad |1+\sqrt{3},1\rangle \times 3\,, \quad |\tfrac{3}{2}+\sqrt{3},\tfrac{1}{2}\rangle \times 8\,, \\   
 &  |\tfrac{3}{2}+\sqrt{3},\tfrac{3}{2}\rangle\,, \quad |2+\sqrt{3},0\rangle \times 6\,, \quad |2+\sqrt{3},1\rangle \times 3 \,, \quad |\tfrac{5}{2}+\sqrt{3},\tfrac{1}{2}\rangle \times 3\,, \quad |3+\sqrt{3},0\rangle \times 1 \,.
\end{split}
\end{equation}
\end{itemize}
The full KK spectrum of massive IIA supergravity for this AdS$_4$ vacuum was recently computed in \cite{Varela:2020wty} using the techniques developed in \cite{Malek:2019eaz,Malek:2020yue}.

\paragraph{\Pu{23795609}} This is an ${\cal N}=1$ vacuum with
\label{P23795609}
\be
V=-\frac{2^8\times 3^{3/2}}{5^{5/2}}\,,
\ee
and $\SU(3)$ continuous symmetry. The solution was first studied in \cite{Guarino:2015qaa} where the four-dimensional bosonic spectrum was computed. The complete four-dimensional mass spectrum including also fermionic fluctuations is
\begin{flalign*}
\qquad\qquad m^2_\psi L^2 &: 4_{\times 1}\,,~ (\tfrac{16}9)_{\times 6}\,,~ 1_{\times 1}\\
m^2_A L^2 &:  6_{\times 1}\,,~(\tfrac{28}9)_{\times 6}\,,~ (\tfrac{25}9)_{\times 6}\,,~2_{\times 1}\,,~(\tfrac49)_{\times 6}\,,~ 0_{\times 8}\,,&\\
m^2_\chi L^2 &: 16_{\times 1}\,,~(\tfrac{64}9)_{\times 6}\,,~6_{\times 2}\,,~(\tfrac{59}{18}+\tfrac{\sqrt{109}}{6})_{\times 6}\,,~4_{\times 1}\,,~(\tfrac{16}9)_{\times 6}\,,&\\
&\quad(\tfrac{59}{18}-\tfrac{\sqrt{109}}{6})_{\times 6}\,,~1_{\times 8}\,,~(\tfrac49)_{\times 12}\,,~0_{\times 8}\,,&\\
m^2_\phi L^2 &: (4+\sqrt{6})_{\times 2}\,,~ (4-\sqrt{6})_{\times 2},~ (\tfrac79)_{\times 6}\,,~ 0_{\times 28}\,,~ (-\tfrac{8}9)_{\times 12}\,,~ (-2)_{\times 8}\,,~ (-\tfrac{20}9)_{\times 12}\,.
\end{flalign*}
The spectrum of low-dimension operators in the dual 3d ${\cal N}=1$ CFT can be organized into the following short
\be
\text{Short}[\tfrac32]_{\times 1}\,,\quad  \text{Short}[\tfrac12]_{\times 8}\,,
\ee
and long multiplets
\be
\begin{split}
&\text{Long}[3,1]_{\times 1}\,,\quad \text{Long}[\tfrac73,1]_{\times 6}\,,\quad \text{Long}\Big[1+\sqrt{\tfrac{109}{36}},\tfrac12\Big]_{\times 6}\,,\\
&\text{Long}\Big[1+\sqrt{6},0\Big]_{\times 2}\,,\quad\text{Long}[2,0]_{\times 8}\,,\quad \text{Long}[\tfrac53,0]_{\times 12}\,.
\end{split}
\ee

\paragraph{\Pu{25697101}} This is an ${\cal N}=1$ vacuum with
\label{P25697101}
\be
V\approx-25.697101\,,
\ee
and $\U(1)$ continuous symmetry. The solution was first studied in \cite{Guarino:2019snw} and the four-dimensional bosonic spectrum was computed in \cite{Guarino:2020jwv}. The complete four-dimensional mass spectrum including also fermionic fluctuations is
\begin{flalign*}
\qquad\qquad m^2_\psi L^2 &: 4.02416_{\times 1}\,,~2.94452_{\times 1}\,,~2.78901_{\times 1}\,,~2.16471_{\times 2}\,,~1.61937_{\times 2}\,,~ 1_{\times 1}\,,~ &\\
m^2_A L^2 &: 6.03020_{\times 1}\,,~5.71869_{\times 2}\,,~4.74884_{\times 2}\,,~4.66048_{\times 1}\,,~4.45905_{\times 1}\,,~ &\\
&\quad 3.63601_{\times 2}\,,~2.89191_{\times 2}\,,~2.01813_{\times 1}\,,~1.91361_{\times 2}\,,~1.22856_{\times 1}\,,~ &\\
&\quad 1.11898_{\times 1}\,,~0.710738_{\times 1}\,,~0.693418_{\times 2}\,,~0.615608_{\times 1}\,,~0.359215_{\times 1}\,,~ &\\
&\quad 0.346824_{\times 2}\,,~0.250555_{\times 2}\,,~0.177244_{\times 2}\,,~0_{\times 1}\,,~ &\\
m^2_\chi L^2 &: 16.0967_{\times 1}\,,~11.7781_{\times 1}\,,~11.1561_{\times 1}\,,~8.66178_{\times 2}\,,~8.65886_{\times 2}\,,~ &\\
&\quad 7.48465_{\times 2}\,,~7.43727_{\times 1}\,,~7.38170_{\times 2}\,,~6.47747_{\times 2}\,,~4.17826_{\times 1}\,,~ &\\
&\quad 4.02416_{\times 1}\,,~3.88453_{\times 2}\,,~3.77560_{\times 2}\,,~3.22624_{\times 1}\,,~3.01303_{\times 2}\,,~ &\\
&\quad 2.94452_{\times 1}\,,~2.78901_{\times 1}\,,~2.19091_{\times 1}\,,~2.16471_{\times 2}\,,~2.04599_{\times 1}\,,~ &\\
&\quad 1.63974_{\times 1}\,,~1.61937_{\times 2}\,,~1.54207_{\times 1}\,,~1.45805_{\times 2}\,,~1.33088_{\times 2}\,,~ &\\
&\quad 0.942689_{\times 2}\,,~0.230565_{\times 1}\,,~0.185227_{\times 1}\,,~0.158783_{\times 2}\,,~0.0850812_{\times 1}\,,~ &\\
&\quad 0.0786929_{\times 1}\,,~0.0430559_{\times 2}\,,~0.0335791_{\times 1}\,,~0.0236049_{\times 2}\,,~0.0149976_{\times 2}\,,~ &\\
&\quad 0.0123450_{\times 1}\,,~0.00175250_{\times 2}\,,~0_{\times 1}\,,~ &\\
m^2_\phi L^2 &: 8.16441_{\times 1}\,,~8.09862_{\times 2}\,,~4.22234_{\times 1}\,,~3.71869_{\times 2}\,,~3.02242_{\times 1}\,,~ &\\
&\quad 2.74884_{\times 2}\,,~2.71014_{\times 1}\,,~2.66477_{\times 2}\,,~0.783875_{\times 1}\,,~0.134180_{\times 1}\,,~ &\\
&\quad 0_{\times 27}\,,~-0.0863888_{\times 2}\,,~-0.569930_{\times 1}\,,~-1.28926_{\times 1}\,,~-1.38439_{\times 1}\,,~ &\\
&\quad -1.44274_{\times 2}\,,~-1.62323_{\times 1}\,,~-1.64078_{\times 1}\,,~-1.69973_{\times 1}\,,~-1.74944_{\times 2}\,,~ &\\
&\quad -1.78317_{\times 1}\,,~-1.82276_{\times 2}\,,~-1.86254_{\times 2}\,,~-1.87655_{\times 1}\,,~-1.95638_{\times 2}\,,~ &\\
&\quad -2.04011_{\times 2}\,,~-2.09876_{\times 1}\,,~-2.10747_{\times 2}\,,~-2.14967_{\times 1}\,,~-2.20661_{\times 1}\,,~ &\\
&\quad -2.23969_{\times 2}\,,~ &\end{flalign*}
The spectrum of low-dimension operators in the dual 3d ${\cal N}=1$ CFT can be organized into the following short
\be
\text{Short}[\tfrac32]_{\times 1}\,,\quad  \text{Short}[\tfrac12]_{\times 1}\,,
\ee
and long multiplets
\be
\begin{split}
&\text{Long}[3.00603,1]_{\times 1}\,,\quad \text{Long}[2.71596,1]_{\times 1}\,,\quad \text{Long}[2.67003,1]_{\times 1}\,,\\
&\text{Long}[2.47130,1]_{\times 2}\,,\quad \text{Long}[2.27254,1]_{\times 2}\,,\quad \text{Long}[3.44309,\tfrac12]_{\times 2}\,,\\
&\text{Long}[3.23581,\tfrac12]_{\times 2}\,,\quad \text{Long}[2.47092,\tfrac12]_{\times 2}\,,\quad \text{Long}[1.98017,\tfrac12]_{\times 1}\,,\\
&\text{Long}[1.93038,\tfrac12]_{\times 1}\,,\quad \text{Long}[1.78052,\tfrac12]_{\times 1}\,,\quad \text{Long}[1.70750,\tfrac12]_{\times 2}\,,\\
&\text{Long}[1.65364,\tfrac12]_{\times 2}\,,\quad \text{Long}[3.72714,0]_{\times 1}\,,\quad \text{Long}[3.71693,0]_{\times 2}\,,\\
&\text{Long}[3.04408,0]_{\times 1}\,,\quad \text{Long}[2.79617,0]_{\times 1}\,,\quad \text{Long}[2.24180,0]_{\times 1}\,,\\
&\text{Long}[1.39848,0]_{\times 2}\,,\quad \text{Long}[1.29169,0]_{\times 1}\,,\quad \text{Long}[1.18325,0]_{\times 1}\,,\\
&\text{Long}[1.12246,0]_{\times 2}\,,\quad \text{Long}[1.11111,0]_{\times 1}\,,\quad \text{Long}[1.04186,0]_{\times 2}\,.
\end{split}
\ee

\paragraph{\Pu{35610235}} This is an ${\cal N}=1$ vacuum with
\label{P35610235}
\be
V\approx-35.610235\,,
\ee
and $\U(1)$ continuous symmetry. The solution was first studied in \cite{Guarino:2019snw} and the four-dimensional bosonic spectrum was computed in \cite{Guarino:2020jwv}. The complete four-dimensional mass spectrum including also fermionic fluctuations is
\begin{flalign*}
\qquad\qquad m^2_\psi L^2 &: 4.96968_{\times 1}\,,~4.73233_{\times 1}\,,~3.20491_{\times 2}\,,~2.80058_{\times 1}\,,~2.55832_{\times 2}\,,~ 1_{\times 1}\,,~ &\\
m^2_A L^2 &: 7.19896_{\times 1}\,,~6.90772_{\times 1}\,,~6.76499_{\times 1}\,,~6.24465_{\times 1}\,,~6.05457_{\times 2}\,,~ &\\
&\quad 5.93897_{\times 2}\,,~5.61325_{\times 1}\,,~4.99513_{\times 2}\,,~4.71417_{\times 2}\,,~4.47408_{\times 1}\,,~ &\\
&\quad 4.15780_{\times 2}\,,~2.74040_{\times 1}\,,~2.55694_{\times 1}\,,~1.41468_{\times 2}\,,~1.12709_{\times 1}\,,~ &\\
&\quad 0.958848_{\times 2}\,,~0.654210_{\times 2}\,,~0.384736_{\times 2}\,,~0_{\times 1}\,,~ &\\
m^2_\chi L^2 &: 19.8787_{\times 1}\,,~18.9293_{\times 1}\,,~12.8196_{\times 2}\,,~11.2023_{\times 1}\,,~10.2333_{\times 2}\,,~ &\\
&\quad 9.91358_{\times 1}\,,~9.73539_{\times 1}\,,~9.29311_{\times 1}\,,~9.12144_{\times 1}\,,~9.06547_{\times 2}\,,~ &\\
&\quad 8.92674_{\times 2}\,,~8.83658_{\times 1}\,,~8.53466_{\times 1}\,,~7.56477_{\times 2}\,,~7.44222_{\times 2}\,,~ &\\
&\quad 6.93688_{\times 1}\,,~4.96968_{\times 1}\,,~4.73233_{\times 1}\,,~4.61641_{\times 1}\,,~4.19619_{\times 1}\,,~ &\\
&\quad 4.09060_{\times 2}\,,~4.04368_{\times 2}\,,~4.02032_{\times 1}\,,~3.95121_{\times 2}\,,~3.69183_{\times 1}\,,~ &\\
&\quad 3.20491_{\times 2}\,,~2.98613_{\times 2}\,,~2.80058_{\times 1}\,,~2.55832_{\times 2}\,,~2.10511_{\times 2}\,,~ &\\
&\quad 1.68144_{\times 2}\,,~1.07685_{\times 2}\,,~0.310663_{\times 1}\,,~0.281223_{\times 2}\,,~0.203311_{\times 2}\,,~ &\\
&\quad 0.157113_{\times 1}\,,~0.0880327_{\times 2}\,,~0_{\times 1}\,,~ &\\
m^2_\phi L^2 &: 10.8555_{\times 1}\,,~10.1416_{\times 1}\,,~9.80922_{\times 1}\,,~8.31518_{\times 2}\,,~7.57067_{\times 1}\,,~ &\\
&\quad 4.76499_{\times 1}\,,~4.61523_{\times 1}\,,~4.24465_{\times 1}\,,~4.11312_{\times 2}\,,~4.10127_{\times 1}\,,~ &\\
&\quad 4.05457_{\times 2}\,,~4.02540_{\times 1}\,,~3.93897_{\times 2}\,,~3.86395_{\times 1}\,,~3.61325_{\times 1}\,,~ &\\
&\quad 2.81436_{\times 2}\,,~2.71417_{\times 2}\,,~2.30308_{\times 1}\,,~0.114560_{\times 2}\,,~0.0680744_{\times 2}\,,~ &\\
&\quad 0.0152488_{\times 1}\,,~0_{\times 27}\,,~-1.13197_{\times 1}\,,~-1.18847_{\times 2}\,,~-1.34579_{\times 2}\,,~ &\\
&\quad -1.44651_{\times 1}\,,~-1.61526_{\times 2}\,,~-1.96087_{\times 2}\,,~-2.23926_{\times 1}\,,~-2.24671_{\times 1}\,,~ &\\
&\quad -2.24908_{\times 2}\,,~ &\end{flalign*}

The spectrum of low-dimension operators in the dual 3d ${\cal N}=1$ CFT can be organized into the following short
\be
\text{Short}[\tfrac32]_{\times 1}\,,\quad  \text{Short}[\tfrac12]_{\times 1}\,,
\ee
and long multiplets
\be
\begin{split}
&\text{Long}[3.22928,1]_{\times 1}\,,\quad \text{Long}[3.17539,1]_{\times 1}\,,\quad \text{Long}[2.79023,1]_{\times 2}\,,\\
&\text{Long}[2.67349,1]_{\times 1}\,,\quad \text{Long}[2.59948,1]_{\times 2}\,,\quad \text{Long}[3.64858,\tfrac12]_{\times 1}\,,\\ 
&\text{Long}[3.54846,\tfrac12]_{\times 1}\,,\quad \text{Long}[3.51089,\tfrac12]_{\times 2}\,,\quad \text{Long}[3.48776,\tfrac12]_{\times 2}\,,\\ 
&\text{Long}[3.42141,\tfrac12]_{\times 1}\,,\quad \text{Long}[3.22804,\tfrac12]_{\times 2}\,,\quad \text{Long}[1.95090,\tfrac12]_{\times 2}\,,\\ 
&\text{Long}[1.79670,\tfrac12]_{\times 2}\,,\quad \text{Long}[4.12016,0]_{\times 1}\,,\quad \text{Long}[4.02017,0]_{\times 1}\,,\\ 
&\text{Long}[3.97264,0]_{\times 1}\,,\quad \text{Long}[3.75041,0]_{\times 2}\,,\quad \text{Long}[3.63380,0]_{\times 1}\,,\\
&\text{Long}[3.02252,0]_{\times 2}\,,\quad \text{Long}[3.00507,0]_{\times 1}\,,\quad \text{Long}[2.03771,0]_{\times 2}\,,\\ 
&\text{Long}[1.55737,0]_{\times 1}\,,\quad \text{Long}[1.53030,0]_{\times 2}\,,\quad \text{Long}[1.39637,0]_{\times 1}\,.
\end{split}
\ee
%

\section{The spectra of BF stable non-supersymmetric \texorpdfstring{AdS$_4$}{AdS4} vacua}
\label{sec:nonsusy}

Here we present the spectra of masses for all fields in the 4d gauged supergravity theory around each of the 9 non-supersymmetric but BF stable AdS$_4$ solutions listed in Table~\ref{tableofvacua}. We do not  explicitly list the massless spin-2 mode present in each one of these vacua. We remind the reader that the BF bound for scalars fields in AdS$_4$ is $m^2L^2 \geq -\frac{9}{4}$.

\paragraph{\Pu{23277304}$_2$}
\label{P23277304-2}
\begin{flalign*}
\qquad\qquad m^2_\psi L^2 &: \tfrac{9}{2}_{\times 1}\,,~\tfrac{3}{2}_{\times 7}\,,~ &\\
m^2_A L^2 &: 3_{\times 14}\,,~0_{\times 14}\,,~ &\\
m^2_\chi L^2 &: 18_{\times 1}\,,~6_{\times 7}\,,~\tfrac{9}{2}_{\times 7}\,,~\tfrac{3}{2}_{\times 14}\,,~\tfrac{1}{2}_{\times 27}\,,~ &\\
m^2_\phi L^2 &: 6_{\times 2}\,,~0_{\times 14}\,,~-1_{\times 54}\,,~ &\end{flalign*}

\paragraph{\Pu{23413628}}
\label{P23413628}
\begin{flalign*}
\qquad\qquad m^2_\psi L^2 &: 4.47622_{\times 1}\,,~1.58653_{\times 6}\,,~1.25039_{\times 1}\,,~ &\\
m^2_A L^2 &: 4.37314_{\times 1}\,,~3.19992_{\times 6}\,,~2.79117_{\times 6}\,,~2.49049_{\times 1}\,,~0.110741_{\times 6}\,,~ 0_{\times 8}\,,~ &\\
m^2_\chi L^2 &: 17.9049_{\times 1}\,,~6.34611_{\times 6}\,,~5.71339_{\times 1}\,,~5.00155_{\times 1}\,,~4.54552_{\times 6}\,,~ &\\
&\quad 2.18869_{\times 6}\,,~1.41227_{\times 1}\,,~1.25322_{\times 8}\,,~0.590312_{\times 12}\,,~0.562809_{\times 6}\,,~ &\\
&\quad 0.177942_{\times 8}\,,~ &\\
m^2_\phi L^2 &: 6.22958_{\times 1}\,,~5.90489_{\times 1}\,,~1.12990_{\times 1}\,,~0_{\times 20}\,,~-0.308840_{\times 8}\,,~ &\\
&\quad -0.954193_{\times 12}\,,~-1.08192_{\times 6}\,,~-1.26438_{\times 1}\,,~-1.39571_{\times 8}\,,~-1.58174_{\times 12}\,,~ &\end{flalign*}

\paragraph{\Pu{23456052}}
\label{P23456052}
\begin{flalign*}
\qquad\qquad m^2_\psi L^2 &: 4.34233_{\times 1}\,,~1.71978_{\times 4}\,,~1.32170_{\times 2}\,,~1.28551_{\times 1}\,,~ &\\
m^2_A L^2 &: 4.29451_{\times 1}\,,~3.62600_{\times 2}\,,~3.31200_{\times 4}\,,~2.66757_{\times 2}\,,~2.39744_{\times 4}\,,~ &\\
&\quad 2.29373_{\times 1}\,,~0.125393_{\times 4}\,,~0.0883439_{\times 2}\,,~0.0526758_{\times 4}\,,~0_{\times 4}\,,~ &\\
m^2_\chi L^2 &: 17.3693_{\times 1}\,,~6.87911_{\times 4}\,,~5.57377_{\times 1}\,,~5.42827_{\times 2}\,,~5.28681_{\times 2}\,,~ &\\
&\quad 5.14204_{\times 1}\,,~4.10476_{\times 4}\,,~2.05956_{\times 2}\,,~1.93584_{\times 4}\,,~1.58453_{\times 4}\,,~ &\\
&\quad 1.57432_{\times 1}\,,~1.42881_{\times 1}\,,~1.08509_{\times 2}\,,~0.817300_{\times 3}\,,~0.763508_{\times 2}\,,~ &\\
&\quad 0.636613_{\times 4}\,,~0.361814_{\times 4}\,,~0.339602_{\times 6}\,,~0.249058_{\times 3}\,,~0.237591_{\times 4}\,,~ &\\
&\quad 0.173874_{\times 1}\,,~ &\\
m^2_\phi L^2 &: 6.29251_{\times 1}\,,~5.78023_{\times 1}\,,~1.14647_{\times 1}\,,~0_{\times 24}\,,~-0.0277198_{\times 2}\,,~ &\\
&\quad -0.145118_{\times 1}\,,~-0.825771_{\times 4}\,,~-0.920226_{\times 2}\,,~-0.987543_{\times 1}\,,~-1.13572_{\times 4}\,,~ &\\
&\quad -1.16949_{\times 3}\,,~-1.26471_{\times 6}\,,~-1.36606_{\times 2}\,,~-1.38584_{\times 4}\,,~-1.56985_{\times 3}\,,~ &\\
&\quad -1.58248_{\times 1}\,,~-1.62006_{\times 6}\,,~-1.71379_{\times 4}\,,~ &\end{flalign*}

\paragraph{\Pu{23456098}}
\label{P23456098}
\begin{flalign*}
\qquad\qquad m^2_\psi L^2 &: 4.34248_{\times 1}\,,~1.72262_{\times 4}\,,~1.32252_{\times 1}\,,~1.30762_{\times 1}\,,~1.28725_{\times 1}\,,~ &\\
m^2_A L^2 &: 4.26513_{\times 1}\,,~3.70813_{\times 1}\,,~3.59326_{\times 1}\,,~3.31546_{\times 4}\,,~2.67942_{\times 1}\,,~ &\\
&\quad 2.63558_{\times 1}\,,~2.39041_{\times 4}\,,~2.30724_{\times 1}\,,~0.123236_{\times 4}\,,~0.0948198_{\times 1}\,,~ &\\
&\quad 0.0753444_{\times 1}\,,~0.0562524_{\times 4}\,,~0.00171607_{\times 1}\,,~0_{\times 3}\,,~ &\\
m^2_\chi L^2 &: 17.3699_{\times 1}\,,~6.89048_{\times 4}\,,~5.56531_{\times 1}\,,~5.47446_{\times 1}\,,~5.41899_{\times 1}\,,~ &\\
&\quad 5.29008_{\times 1}\,,~5.23049_{\times 1}\,,~5.14900_{\times 1}\,,~4.09224_{\times 4}\,,~2.10385_{\times 1}\,,~ &\\
&\quad 1.99973_{\times 1}\,,~1.93008_{\times 4}\,,~1.59837_{\times 4}\,,~1.59825_{\times 1}\,,~1.41838_{\times 1}\,,~ &\\
&\quad 1.10589_{\times 1}\,,~1.08716_{\times 1}\,,~0.809770_{\times 3}\,,~0.798969_{\times 1}\,,~0.751003_{\times 1}\,,~ &\\
&\quad 0.638581_{\times 4}\,,~0.356887_{\times 4}\,,~0.348094_{\times 3}\,,~0.317619_{\times 3}\,,~0.254423_{\times 3}\,,~ &\\
&\quad 0.239296_{\times 4}\,,~0.171334_{\times 1}\,,~ &\\
m^2_\phi L^2 &: 6.29845_{\times 1}\,,~5.77263_{\times 1}\,,~1.14238_{\times 1}\,,~0.0395421_{\times 1}\,,~0_{\times 25}\,,~ &\\
&\quad -0.214610_{\times 1}\,,~-0.834473_{\times 4}\,,~-0.874352_{\times 1}\,,~-0.960622_{\times 1}\,,~-0.977451_{\times 1}\,,~ &\\
&\quad -1.13165_{\times 4}\,,~-1.20232_{\times 3}\,,~-1.27193_{\times 3}\,,~-1.28289_{\times 3}\,,~-1.34552_{\times 1}\,,~ &\\
&\quad -1.35168_{\times 1}\,,~-1.37196_{\times 4}\,,~-1.56255_{\times 3}\,,~-1.57899_{\times 3}\,,~-1.58451_{\times 1}\,,~ &\\
&\quad -1.64152_{\times 3}\,,~-1.71663_{\times 4}\,,~ &\end{flalign*}

\paragraph{\Pu{23456778}}
\label{P23456778}
\begin{flalign*}
\qquad\qquad m^2_\psi L^2 &: 4.35284_{\times 1}\,,~1.59271_{\times 6}\,,~1.27532_{\times 1}\,,~ &\\
m^2_A L^2 &: 4.67676_{\times 1}\,,~3.18378_{\times 6}\,,~2.71517_{\times 6}\,,~2.13568_{\times 1}\,,~0.150049_{\times 6}\,,~  0_{\times 8}\,,~ &\\
m^2_\chi L^2 &: 17.4114_{\times 1}\,,~6.37083_{\times 6}\,,~5.68294_{\times 1}\,,~5.10127_{\times 1}\,,~4.63416_{\times 6}\,,~ &\\
&\quad 2.06315_{\times 6}\,,~1.57957_{\times 1}\,,~1.21768_{\times 8}\,,~0.566626_{\times 12}\,,~0.509940_{\times 6}\,,~ &\\
&\quad 0.191836_{\times 8}\,,~ &\\
m^2_\phi L^2 &: 6.21445_{\times 1}\,,~5.92510_{\times 1}\,,~1.14480_{\times 1}\,,~0_{\times 20}\,,~-0.158522_{\times 8}\,,~ &\\
&\quad -0.859766_{\times 12}\,,~-1.06147_{\times 6}\,,~-1.28435_{\times 1}\,,~-1.62283_{\times 8}\,,~-1.70679_{\times 12}\,,~ &\end{flalign*}

\paragraph{\Pu{23458779}}
\label{P23458779}
\begin{flalign*}
\qquad\qquad m^2_\psi L^2 &: 4.35028_{\times 1}\,,~1.65919_{\times 4}\,,~1.45670_{\times 2}\,,~1.27671_{\times 1}\,,~ &\\
m^2_A L^2 &: 4.59692_{\times 1}\,,~3.26745_{\times 2}\,,~3.24580_{\times 4}\,,~2.82061_{\times 2}\,,~2.55925_{\times 4}\,,~ &\\
&\quad 2.16737_{\times 1}\,,~0.147874_{\times 4}\,,~0.132118_{\times 2}\,,~0.0115978_{\times 4}\,,~0_{\times 4}\,,~ &\\
m^2_\chi L^2 &: 17.4011_{\times 1}\,,~6.63676_{\times 4}\,,~5.82678_{\times 2}\,,~5.65794_{\times 1}\,,~5.10682_{\times 1}\,,~ &\\
&\quad 5.07034_{\times 2}\,,~4.37936_{\times 4}\,,~2.09910_{\times 2}\,,~2.03236_{\times 4}\,,~1.54604_{\times 1}\,,~ &\\
&\quad 1.43198_{\times 1}\,,~1.36513_{\times 4}\,,~0.997243_{\times 3}\,,~0.796384_{\times 2}\,,~0.614906_{\times 2}\,,~ &\\
&\quad 0.604950_{\times 4}\,,~0.458648_{\times 6}\,,~0.452034_{\times 4}\,,~0.202040_{\times 4}\,,~0.200635_{\times 3}\,,~ &\\
&\quad 0.193346_{\times 1}\,,~ &\\
m^2_\phi L^2 &: 6.22349_{\times 1}\,,~5.89550_{\times 1}\,,~1.16111_{\times 1}\,,~0.0852111_{\times 1}\,,~0_{\times 24}\,,~ &\\
&\quad -0.480612_{\times 2}\,,~-0.627003_{\times 3}\,,~-0.804699_{\times 4}\,,~-0.935656_{\times 2}\,,~-1.04651_{\times 6}\,,~ &\\
&\quad -1.12566_{\times 4}\,,~-1.22285_{\times 1}\,,~-1.55930_{\times 4}\,,~-1.58649_{\times 1}\,,~-1.59719_{\times 2}\,,~ &\\
&\quad -1.63609_{\times 3}\,,~-1.69541_{\times 4}\,,~-1.70677_{\times 6}\,,~ &\end{flalign*}

\paragraph{\Pu{23512689}}
\label{P23512689}
\begin{flalign*}
\qquad\qquad m^2_\psi L^2 &: 4.11184_{\times 1}\,,~1.87329_{\times 4}\,,~1.15576_{\times 3}\,,~ &\\
m^2_A L^2 &: 4.15307_{\times 3}\,,~3.45088_{\times 4}\,,~2.28714_{\times 3}\,,~1.94511_{\times 4}\,,~0.191140_{\times 8}\,,~  0_{\times 6}\,,~ &\\
m^2_\chi L^2 &: 16.4474_{\times 1}\,,~7.49316_{\times 4}\,,~5.99779_{\times 3}\,,~4.62306_{\times 3}\,,~3.65501_{\times 4}\,,~ &\\
&\quad 1.76841_{\times 8}\,,~1.67139_{\times 5}\,,~1.35101_{\times 3}\,,~0.736915_{\times 4}\,,~0.456425_{\times 3}\,,~ &\\
&\quad 0.234629_{\times 8}\,,~0.178226_{\times 9}\,,~0.0434607_{\times 1}\,,~ &\\
m^2_\phi L^2 &: 6.72740_{\times 1}\,,~5.28662_{\times 1}\,,~0.629766_{\times 5}\,,~0.584358_{\times 1}\,,~0_{\times 22}\,,~ &\\
&\quad -0.729624_{\times 4}\,,~-0.982712_{\times 5}\,,~-1.17591_{\times 8}\,,~-1.58552_{\times 1}\,,~-1.58816_{\times 9}\,,~ &\\
&\quad -1.75110_{\times 9}\,,~-1.96422_{\times 4}\,,~ &\end{flalign*}

\paragraph{\Pu{23715872}}
\label{P23715872}
\begin{flalign*}
\qquad\qquad m^2_\psi L^2 &: 4.05937_{\times 1}\,,~1.97687_{\times 2}\,,~1.82407_{\times 1}\,,~1.81894_{\times 1}\,,~1.37617_{\times 2}\,,~ &\\
&\quad 1.06398_{\times 1}\,,~ &\\
m^2_A L^2 &: 5.48592_{\times 1}\,,~3.72565_{\times 2}\,,~3.37821_{\times 2}\,,~3.31705_{\times 1}\,,~3.21520_{\times 1}\,,~ &\\
&\quad 2.87561_{\times 1}\,,~2.71437_{\times 2}\,,~2.33181_{\times 1}\,,~1.91237_{\times 2}\,,~1.70402_{\times 1}\,,~ &\\
&\quad 0.473109_{\times 1}\,,~0.451036_{\times 2}\,,~0.417981_{\times 1}\,,~0.159393_{\times 2}\,,~0.0756985_{\times 2}\,,~ &\\
&\quad 0.0671494_{\times 2}\,,~0.0190147_{\times 1}\,,~0.0135513_{\times 2}\,,~0_{\times 1}\,,~ &\\
m^2_\chi L^2 &: 16.2375_{\times 1}\,,~7.90749_{\times 2}\,,~7.29628_{\times 1}\,,~7.27574_{\times 1}\,,~6.18256_{\times 1}\,,~ &\\
&\quad 5.89581_{\times 2}\,,~5.50467_{\times 2}\,,~5.49317_{\times 1}\,,~4.78789_{\times 1}\,,~4.25592_{\times 1}\,,~ &\\
&\quad 3.87092_{\times 2}\,,~3.00895_{\times 1}\,,~2.47563_{\times 2}\,,~2.03969_{\times 1}\,,~1.89184_{\times 2}\,,~ &\\
&\quad 1.82134_{\times 1}\,,~1.71997_{\times 1}\,,~1.64628_{\times 2}\,,~1.29125_{\times 2}\,,~1.21935_{\times 2}\,,~ &\\
&\quad 1.20274_{\times 1}\,,~0.983986_{\times 2}\,,~0.938584_{\times 2}\,,~0.893911_{\times 1}\,,~0.587593_{\times 2}\,,~ &\\
&\quad 0.562868_{\times 1}\,,~0.439865_{\times 1}\,,~0.371783_{\times 1}\,,~0.346838_{\times 2}\,,~0.303637_{\times 1}\,,~ &\\
&\quad 0.294338_{\times 1}\,,~0.262899_{\times 2}\,,~0.236847_{\times 2}\,,~0.0339378_{\times 2}\,,~0.0286117_{\times 2}\,,~ &\\
&\quad 0.0205378_{\times 1}\,,~0.0192755_{\times 2}\,,~0.0114180_{\times 1}\,,~ &\\
m^2_\phi L^2 &: 6.67265_{\times 1}\,,~5.70582_{\times 1}\,,~1.71484_{\times 2}\,,~1.67033_{\times 1}\,,~1.37757_{\times 1}\,,~ &\\
&\quad 0.666216_{\times 1}\,,~0_{\times 27}\,,~-0.0891377_{\times 2}\,,~-0.547035_{\times 1}\,,~-0.833421_{\times 2}\,,~ &\\
&\quad -0.907516_{\times 1}\,,~-0.915646_{\times 1}\,,~-1.14803_{\times 1}\,,~-1.17006_{\times 2}\,,~-1.29452_{\times 1}\,,~ &\\
&\quad -1.30728_{\times 1}\,,~-1.35666_{\times 2}\,,~-1.39437_{\times 2}\,,~-1.64003_{\times 2}\,,~-1.70815_{\times 2}\,,~ &\\
&\quad -1.74308_{\times 1}\,,~-1.87815_{\times 2}\,,~-1.96526_{\times 1}\,,~-1.96573_{\times 2}\,,~-2.06671_{\times 1}\,,~ &\\
&\quad -2.06900_{\times 2}\,,~-2.06965_{\times 1}\,,~-2.09083_{\times 1}\,,~-2.09758_{\times 2}\,,~-2.13338_{\times 2}\,,~ &\\
&\quad -2.18141_{\times 1}\,,~ &\end{flalign*}

\paragraph{\Pu{355983403}}
\label{P355983403}
\begin{flalign*}
\qquad\qquad m^2_\psi L^2 &: 5.03505_{\times 1}\,,~4.90092_{\times 1}\,,~3.25015_{\times 1}\,,~3.14370_{\times 1}\,,~2.75155_{\times 1}\,,~ &\\
&\quad 2.57142_{\times 1}\,,~2.41902_{\times 1}\,,~1.00004_{\times 1}\,,~ &\\
m^2_A L^2 &: 7.25938_{\times 1}\,,~7.20263_{\times 1}\,,~7.10141_{\times 1}\,,~6.95967_{\times 1}\,,~6.35969_{\times 1}\,,~ &\\
&\quad 5.98220_{\times 1}\,,~5.78977_{\times 1}\,,~5.68272_{\times 1}\,,~5.22257_{\times 1}\,,~5.06055_{\times 1}\,,~ &\\
&\quad 4.92408_{\times 1}\,,~4.73720_{\times 1}\,,~4.40682_{\times 1}\,,~4.27011_{\times 1}\,,~4.17636_{\times 1}\,,~ &\\
&\quad 3.97613_{\times 1}\,,~2.79003_{\times 1}\,,~2.68509_{\times 1}\,,~1.44360_{\times 1}\,,~1.36673_{\times 1}\,,~ &\\
&\quad 1.09360_{\times 1}\,,~0.968181_{\times 1}\,,~0.863857_{\times 1}\,,~0.673824_{\times 1}\,,~0.670555_{\times 1}\,,~ &\\
&\quad 0.394090_{\times 1}\,,~0.367021_{\times 1}\,,~0.00324705_{\times 1}\,,~ &\\
m^2_\chi L^2 &: 20.1402_{\times 1}\,,~19.6037_{\times 1}\,,~13.0006_{\times 1}\,,~12.5748_{\times 1}\,,~11.0062_{\times 1}\,,~ &\\
&\quad 10.4291_{\times 1}\,,~10.2857_{\times 1}\,,~10.1418_{\times 1}\,,~9.87119_{\times 1}\,,~9.67608_{\times 1}\,,~ &\\
&\quad 9.42621_{\times 1}\,,~9.41722_{\times 1}\,,~8.96914_{\times 1}\,,~8.74559_{\times 1}\,,~8.74471_{\times 1}\,,~ &\\
&\quad 8.60848_{\times 1}\,,~8.05204_{\times 1}\,,~7.63541_{\times 1}\,,~7.46906_{\times 1}\,,~7.37218_{\times 1}\,,~ &\\
&\quad 7.00906_{\times 1}\,,~6.89305_{\times 1}\,,~5.00512_{\times 1}\,,~4.97228_{\times 1}\,,~4.87874_{\times 1}\,,~ &\\
&\quad 4.77513_{\times 1}\,,~4.28864_{\times 1}\,,~4.25789_{\times 1}\,,~4.23611_{\times 1}\,,~4.00016_{\times 1}\,,~ &\\
&\quad 3.98591_{\times 1}\,,~3.83188_{\times 1}\,,~3.74524_{\times 1}\,,~3.38312_{\times 1}\,,~3.25582_{\times 1}\,,~ &\\
&\quad 3.14876_{\times 1}\,,~3.00171_{\times 1}\,,~2.74757_{\times 1}\,,~2.64195_{\times 1}\,,~2.57394_{\times 1}\,,~ &\\
&\quad 2.42190_{\times 1}\,,~2.13557_{\times 1}\,,~2.13069_{\times 1}\,,~1.69644_{\times 1}\,,~1.65198_{\times 1}\,,~ &\\
&\quad 1.00783_{\times 1}\,,~0.867934_{\times 1}\,,~0.286209_{\times 1}\,,~0.276690_{\times 1}\,,~0.275490_{\times 1}\,,~ &\\
&\quad 0.216605_{\times 1}\,,~0.214997_{\times 1}\,,~0.138143_{\times 1}\,,~0.0910165_{\times 1}\,,~0.0803230_{\times 1}\,,~ &\\
&\quad 0.0000126900_{\times 1}\,,~ &\\
m^2_\phi L^2 &: 11.0147_{\times 1}\,,~10.4898_{\times 1}\,,~9.70105_{\times 1}\,,~8.39197_{\times 1}\,,~8.08150_{\times 1}\,,~ &\\
&\quad 7.65928_{\times 1}\,,~5.19712_{\times 1}\,,~4.95637_{\times 1}\,,~4.72816_{\times 1}\,,~4.35072_{\times 1}\,,~ &\\
&\quad 4.34468_{\times 1}\,,~4.31247_{\times 1}\,,~4.28617_{\times 1}\,,~4.00337_{\times 1}\,,~3.96357_{\times 1}\,,~ &\\
&\quad 3.79000_{\times 1}\,,~3.78372_{\times 1}\,,~3.65945_{\times 1}\,,~3.20239_{\times 1}\,,~2.87878_{\times 1}\,,~ &\\
&\quad 2.72982_{\times 1}\,,~2.66281_{\times 1}\,,~2.35844_{\times 1}\,,~2.25944_{\times 1}\,,~0.203476_{\times 1}\,,~ &\\
&\quad 0.186242_{\times 1}\,,~0_{\times 28}\,,~-0.00336031_{\times 1}\,,~-0.201939_{\times 1}\,,~-1.18509_{\times 1}\,,~ &\\
&\quad -1.19989_{\times 1}\,,~-1.20241_{\times 1}\,,~-1.31722_{\times 1}\,,~-1.32040_{\times 1}\,,~-1.48307_{\times 1}\,,~ &\\
&\quad -1.60717_{\times 1}\,,~-1.63632_{\times 1}\,,~-1.99407_{\times 1}\,,~-2.06215_{\times 1}\,,~-2.24013_{\times 1}\,,~ &\\
&\quad -2.24248_{\times 1}\,,~-2.24659_{\times 1}\,,~-2.24727_{\times 1}\,,~ &\end{flalign*}


\bibliography{references}
\bibliographystyle{JHEP}

\end{document}